\pdfoutput=1
\documentclass[aps,prd,onecolumn,superscriptaddress,nofootinbib, a4paper]{revtex4}
\usepackage[all]{xy}
\usepackage{graphicx}
\usepackage{slashed}
\usepackage{amsmath,bbm,latexsym,amssymb}
\usepackage{epsfig}
\usepackage{epstopdf}
\usepackage{float} 
\usepackage{tensor}
\usepackage{booktabs}
\usepackage{multirow}
\usepackage{cancel}

\def\ifmath#1{\relax\ifmmode #1\else $#1$\fi}

\newcommand{\be}{\begin{equation}}
\newcommand{\ee}{\end{equation}}
\newcommand{\ba}{\begin{eqnarray}}
\newcommand{\ea}{\end{eqnarray}}

\def\eqs#1#2{Eqs.~(\ref{#1}) and (\ref{#2})}

\newcommand{\Z}[1]{\ensuremath{\mathbbm{Z}_{#1}}} 

\usepackage{xparse}
\usepackage[hyperfootnotes=false]{hyperref}
\hypersetup{
    colorlinks=true,
    linkcolor=blue,
    filecolor=blue,      
    urlcolor=blue,
    citecolor=blue
}

\makeatletter
\renewcommand*\env@matrix[1][\arraystretch]{%
  \edef\arraystretch{#1}%
  \hskip -\arraycolsep
  \let\@ifnextchar\new@ifnextchar
  \array{*\c@MaxMatrixCols c}}
\makeatother

\usepackage{tabularx}
\newcolumntype{C}[1]{>{\centering\arraybackslash}p{#1}}

\begin{document}
\vspace*{1cm}


\title{ 
Novel two component dark matter features\\ 
in the $Z_2 \times Z_2$ 3HDM} 

\author{Rafael Boto}%
\email{rafael.boto@tecnico.ulisboa.pt}
\affiliation{Departamento de
	F\'isica and CFTP,  Instituto Superior T\'ecnico,\\  \small\em
	Universidade de Lisboa, Av
	Rovisco Pais, 1, P-1049-001 Lisboa, Portugal}
\author{Pedro N. Figueiredo}%
\email{pedro.m.figueiredo@tecnico.ulisboa.pt}
\affiliation{Departamento de
	F\'isica and CFTP,  Instituto Superior T\'ecnico,\\  \small\em
	Universidade de Lisboa, Av
	Rovisco Pais, 1, P-1049-001 Lisboa, Portugal}
\author{Jorge C. Romão}%
\email{jorge.romao@tecnico.ulisboa.pt}
\affiliation{Departamento de
	F\'isica and CFTP,  Instituto Superior T\'ecnico,\\  \small\em
	Universidade de Lisboa, Av
	Rovisco Pais, 1, P-1049-001 Lisboa, Portugal}
\author{Jo\~ao P.\ Silva}%
\email{jpsilva@cftp.ist.utl.pt}
\affiliation{Departamento de
	F\'isica and CFTP,  Instituto Superior T\'ecnico,\\  \small\em
	Universidade de Lisboa, Av
	Rovisco Pais, 1, P-1049-001 Lisboa, Portugal}

\begin{abstract}
We discuss the constraints and phenomenology of the
$\Z2\times\Z2$ three Higgs doublet model (3HDM) with two inert scalars, originating
two dark matter (DM) particles.
We elucidate the competing vacua and we submit the model
to all theoretical and current experimental constraints.
We find unexplored regions of parameter space  and investigate their experimental signatures.
In particular, we find regions where the two DM particles
contribute equally to the relic DM density.
\end{abstract}

\maketitle

\section{Introduction}
\label{sec:intro}

The quest to understand the nature of the elusive dark matter (DM)
in the context of a complete model of particle physics persists as
one of the paramount challenges in modern physics.
It is embarrassing that a very precise Standard Model (SM) has
been developed in order to explain around 15\% of matter in the Universe,
while the 85\% DM remains unexplained \cite{Planck:2013oqw}.
On the other hand,
the SM postulates that there is one fundamental scalar doublet
\cite{Englert:1964et,Higgs:1964pj},
predicting that there will be a massive neutral scalar particle; the Higgs boson ($h$).
This was confirmed at LHC in 2012 with the discovery of the first
neutral scalar at $m_h=$125GeV \cite{ATLAS:2012yve,CMS:2012qbp}.
But having one single doublet is the simplest, but an otherwise arbitrary imposition.
Indeed, the number of scalars (as the number of fermion families
a few decades ago) must be determined experimentally.
And, the so-called N Higgs doublet models (NHDM) also offer an avenue
to solve the DM problem.

In its simplest form, one would add one single scalar doublet to the SM,
odd under a $\Z2$ symmetry that leaves all the SM fields unchanged.
If, upon spontaneous symmetry breaking,
the new field does not develop a vacuum expectation value, then the
$\Z2$ symmetry remains unbroken and will be reflected in the particle spectrum.
The lightest of the $\Z2$-odd particles cannot decay and, if it is neutral,
it is a candidate for DM; this is the so-called Inert Doublet Model (IDM)
\cite{Deshpande:1977rw,Ma:2006km,Barbieri:2006dq,LopezHonorez:2006gr}.
There are many interesting articles on the IDM,
including, for example
\cite{Ilnicka:2015jba,Belyaev:2016lok,Kalinowski:2020rmb,Datta:2016nfz}.
The upshot of all theoretical, collider, astrophysical and cosmological constraints
is that the DM candidate can only have its mass restricted to two
regions; one region around $m_h/2$ and another region with mass above
around 500GeV.
The exclusion of the intermediate region comes from an interplay between the requirements
from relic density and the constraints from direct detection (DD) experiments.
In addition, a variety of indirect detection (ID) constraints can arise
from DM annihilation into
photons \cite{Modak:2015uda,Queiroz:2015utg,Garcia-Cely:2015khw}
cosmic rays \cite{Nezri:2009jd},
or neutrinos \cite{Agrawal:2008xz,Andreas:2009hj}.

The DM (intermediate) mass region can be extended in theories
that have more than one DM component \cite{Boehm:2003ha};
the so-called multi-component DM models
\cite{Zurek:2008qg,Batell:2009vb,Profumo:2009tb}.
This is due to the possibility that the various components contribute to the
relic density and also due to the new processes of
co-annihilation \cite{Mizuta:1992qp,Ellis:1998kh},
DM conversion between sectors \cite{Liu:2011aa},
and/or semi-annihilation \cite{DEramo:2010keq}.
Examples involving exclusively new scalars include
models based on $\Z4$
\cite{Belanger:2012vp,Belanger:2021lwd}
or $\Z2\times\Z2$
\cite{Belanger:2011ww,Aoki:2012ub,Bhattacharya:2016ysw,Hernandez-Sanchez:2020aop}.
An interesting feature of multi-component DM models is that they can be used
to explain putative anomalies in uncorrelated DM signals which require different
DM mass scales.
And they may yield signal detectable at the FCC; see for example
\cite{Bhattacharya:2022wtr,Bhattacharya:2022qck}.

In this article, we study a three Higgs doublet model (3HDM),
with a $\Z2\times\Z2$ symmetry, where two scalar doublets are inert.
This leads to two separate DM sectors and two natural DM candidates.
Whilst in models based on $\Z4$
\cite{Belanger:2012vp,Belanger:2021lwd} the existence of two DM candidates
hinges on a suitable choice of masses such that decays of one sector
into the other is kinematically forbidden, this is not the case here,
where such decays are symmetry forbidden.

We introduce the $\Z2\times\Z2$ potential in section~\ref{sec:pot}, where we also discuss
the bounded from below (BFB) conditions.
In sections~\ref{sec:vacua} and \ref{sec:mineq},
we discuss in detail all possible vacua, and the conditions guaranteeing that
the double inert vacuum is indeed the global minimum,
improving on the conditions in \cite{Hernandez-Sanchez:2020aop}.
In sections~\ref{sec:scan} and \ref{sec:further} we set up the scan and
list all collider, astrophysical and cosmological constraints which we
will use in our simulations.
This model has a number of interesting processes relevant for DM studies,
including DM conversion and co-annihilation, which we discuss in
section~\ref{sec:processes}.
The results of our scan, and their discussion and implications are presented
in section~\ref{sec:results}.
We outline our conclusions in section~\ref{sec:concl}.
For completeness,
we include in appendix~\ref{app:masses} the formulas for the scalar masses
in the various vacua,
while appendix~\ref{app:f7f10_cond} contains some derivations concerning the
minimization of the angular part.

\section{The \texorpdfstring{$\Z2\times\Z2$}{Z2xZ2} potential}
\label{sec:pot}

\subsection{Notation}
\label{subsec:notation}

Consider a three Higgs doublet model (3HDM) with the
$\Z2\times\Z2$ symmetry\footnote{We will only use the
notation $\Z2^\prime$ in cases where the distinction becomes necessary.}
\begin{align}
 \label{eq:20}
 \Z2 :&\ \phi_1 \to -\phi_1,\quad \phi_2\to\phi_2,\quad
  \phi_3\to  \phi_3\, ,  \\
  \Z2' :&\ \phi_1 \to \phi_1,\quad\phi_2 \to -\phi_2,\quad
  \phi_3 \to \phi_3\, .
\end{align}
The quadratic part of the potential is
\be
V_2 = m_{11}^2\phi_1^\dagger\phi_1 + m_{22}^2\phi_2^\dagger\phi_2 + m_{33}^2\phi_3^\dagger\phi_3
\, ,
\label{Z2Z2quadratic}
\ee
while its quartic part reads \cite{Boto:2022uwv}
\begin{align}
V_{4}=&
\lambda_1(\phi_1^\dagger\phi_1)^2
+\lambda_2(\phi_2^\dagger\phi_2)^2
+\lambda_3(\phi_3^\dagger\phi_3)^2+
\lambda_4(\phi_1^\dagger\phi_1)(\phi_2^\dagger\phi_2)
+\lambda_5(\phi_1^\dagger\phi_1)(\phi_3^\dagger\phi_3)\nonumber\\[8pt]
& 
+\lambda_6(\phi_2^\dagger\phi_2)(\phi_3^\dagger\phi_3)
+\lambda_7(\phi_1^\dagger\phi_2)(\phi_2^\dagger\phi_1)
+\lambda_8(\phi_1^\dagger\phi_3)(\phi_3^\dagger\phi_1)
+\lambda_9(\phi_2^\dagger\phi_3)(\phi_3^\dagger\phi_2)\nonumber\\[8pt]
&
+\left[\lambda''_{10}(\phi_1^\dagger\phi_2)^2 +
  \lambda''_{11}(\phi_1^\dagger\phi_3)^2 +
  \lambda''_{12}(\phi_2^\dagger\phi_3)^2
  +
\text{h.c.}\right]\, .
\label{Z2Z2quartic}
\end{align}

An alternative notation for the general $\Z2\times\Z2$ symmetric
3HDM potential, used in \cite{Hernandez-Sanchez:2020aop,Hernandez-Sanchez:2022dnn},
has the following form \cite{Ivanov:2011ae,Keus:2013hya}:  
\ba
\label{potentialmoretti}
V &=& V_0+V_{\Z2\times\Z2},\\[1mm]
V_0 &=& - \mu^2_{1} (\phi_1^\dagger \phi_1) -\mu^2_2 (\phi_2^\dagger \phi_2) - \mu^2_3(\phi_3^\dagger \phi_3) \nonumber
+ \lambda_{11} (\phi_1^\dagger \phi_1)^2+ \lambda_{22} (\phi_2^\dagger \phi_2)^2  + \lambda_{33} (\phi_3^\dagger \phi_3)^2 \nonumber\\
&& + \lambda_{12}  (\phi_1^\dagger \phi_1)(\phi_2^\dagger \phi_2)
 + \lambda_{23}  (\phi_2^\dagger \phi_2)(\phi_3^\dagger \phi_3) + \lambda_{31} (\phi_3^\dagger \phi_3)(\phi_1^\dagger \phi_1) \nonumber\\
&& + \lambda'_{12} (\phi_1^\dagger \phi_2)(\phi_2^\dagger \phi_1) 
 + \lambda'_{23} (\phi_2^\dagger \phi_3)(\phi_3^\dagger \phi_2) + \lambda'_{31} (\phi_3^\dagger \phi_1)(\phi_1^\dagger \phi_3),  \nonumber\\[1mm]
V_{\Z2\times\Z2}&=&  \lambda_1 (\phi_1^\dagger \phi_2)^2 + \lambda_2(\phi_2^\dagger \phi_3)^2 + \lambda_3 (\phi_3^\dagger \phi_1)^2 + \mathrm{h.c.}\, .
\nonumber 
\ea
Comparing with \eqs{Z2Z2quadratic}{Z2Z2quartic}, the relation between the two notations is
\begin{align}
  \label{eq:lambasmoretti}
&  - \mu^2_{1} \to m^2_{11},\
   - \mu^2_{2} \to m^2_{22},\
    - \mu^2_{3} \to m^2_{33},\\[+2mm]
 & \lambda_{11}\to  \lambda_1,\
  \lambda_{22}\to  \lambda_2,\
  \lambda_{33}\to  \lambda_3,\
  \lambda_{12}\to \lambda_4,\
  \lambda_{31}\to \lambda_5,\\[+2mm]
 & \lambda_{23}\to \lambda_6,\ 
  \lambda'_{12}\to \lambda_7,\
  \lambda'_{31}\to \lambda_8,\
  \lambda'_{23}\to \lambda_9 \\[+2mm]
& \lambda_1\to\lambda''_{10},\
  \lambda_3\to\lambda''_{11},\
 \lambda_2\to \lambda''_{12}  \, .
\end{align}


\subsection{Bounded from below and unitarity conditions}
\label{subsec:theorcond}

Any physically meaningful potential must be bounded from below (BFB).
It is quite interesting that,
although the $\Z2\times\Z2$ 3HDM was first proposed by Weinberg in
1976 \cite{Weinberg:1976hu},
there is still no known complete BFB necessary and sufficient conditions for this model.
This is a testament to the intricacies involved in assessing the properties
of a potential, including its vacua structure and BFB conditions.

When one has only one Higgs doublet, the possible vacua will unavoidably
leave a remnant $U(1)$ gauge symmetry, corresponding to a
massless photon.
In contrast,
when there are two or more Higgs doublets,
the list of possible vacua always includes cases where there
is a massless photon - the so-called neutral vacua -
and cases where there is no remnant $U(1)$ gauge symmetry, corresponding to a
massive ``photon'' - which are dubbed charge breaking (CB) vacua.
By a conceptual extension,
one classifies the ``directions'' in field space (now, not necessarily
solutions of the stationarity equations) as neutral and CB, respectively.
And, the BFB conditions along such directions are classified as
BFB-n and BFB-c, respectively.

The necessary and sufficient conditions for the $\Z2\times\Z2$ 3HDM
potential to be BFB-n conditions were present in \cite{Grzadkowski:2009bt}.
They find
\begin{align}
&
  \lambda_1 >0,\ \lambda_2 > 0,\ \lambda_3 >0\, ,
\label{eq:19a3}\\[+2mm]
&
  \lambda_x > -2\sqrt{\lambda_1\lambda_2},
  \ \lambda_y > -2 \sqrt{\lambda_1\lambda_3},\
  \lambda_z >-2 \sqrt{\lambda_2\lambda_3}\, ,
\label{eq:19b3}\\[+2mm]
&
  \left\{
    \lambda_x \sqrt{\lambda_3} +\lambda_y \sqrt{\lambda_2}
    +\lambda_z \sqrt{\lambda_1} \ge 0
  \right\}
  \cup
  \left\{
  \lambda_1
  \lambda_z^2+\lambda_2\lambda_y^2+\lambda_3\lambda_x^2
  -4\lambda_1\lambda_2\lambda_3  
  - \lambda_x\lambda_y\lambda_z < 0  
  \right\}\, ,
\label{eq:19c3}
\end{align}
where
\begin{eqnarray}
  \lambda_x &=& \lambda_4+\text{min}(0,\lambda_7-2|\lambda''_{10}|)\, ,
\nonumber\\
  \lambda_y &=& \lambda_5+\text{min}(0,\lambda_8-2|\lambda''_{11}|)\, ,
\nonumber\\
  \lambda_z &=& \lambda_6+\text{min}(0,\lambda_9-2|\lambda''_{12}|)\, .
\label{eq:13d}
\end{eqnarray}
Later, Ref.~\cite{Faro:2019vcd} showed that the BFB-c conditions in
\cite{Grzadkowski:2009bt} were sufficient but not necessary.
Given the lack of necessary and sufficient conditions
for BFB common to many 3HDM,
Ref.~\cite{Boto:2022uwv} introduced a general method to obtain
sufficient conditions for BFB-n and for BFB-c in any model.
Here, we use the  necessary and sufficient BFB-n conditions from
\cite{Grzadkowski:2009bt} together with the sufficient
BFB-c conditions from \cite{Boto:2022uwv}.

%
We use the perturbative unitarity conditions for the $\Z2\times\Z2$ 3HDM
found in \cite{Moretti:2015cwa}
and in \cite{Bento:2022vsb}, which generalizes into complex coefficients and
contains also the unitarity conditions for all
other symmetry-constrained 3HDM.

\section{The possible vacua}
\label{sec:vacua}

This section is devoted to the identification of
the possible vacua of the 3HDM $\Z2\times\Z2$ model
and the criteria ensuring that the inert vacuum corresponds to the global minimum.
We start by describing the various vacua.
Then, we explain how numerical and analytical explorations show that 
not all vacua are contained in Ref.~\cite{Hernandez-Sanchez:2020aop}.

\subsection{Neutral vacua}
\label{subsec:neutral-vacua}

The most general neutral vacuum configuration may be parametrized as
\begin{equation}
  \label{eq:1n}
  \langle \phi_1 \rangle  =
  \begin{pmatrix}
    0\\
    v_1 e^{i\xi_1}
  \end{pmatrix},\quad
  \langle \phi_2 \rangle = 
  \begin{pmatrix}
    0\\
    v_2 e^{i\xi_2}
  \end{pmatrix},\quad
  \langle \phi_3 \rangle =
  \begin{pmatrix}
    0\\
    v_3
  \end{pmatrix} .
\end{equation}
Various of its distinct incarnations were studied in Ref.~\cite{Hernandez-Sanchez:2020aop}.
We follow their notation for the classification, which is shown in
the upper part of Table~\ref{tab:1}.
\begin{table}[htb]
  \centering
  \begin{tabular}{|c|c|c|c|}\hline
    Name &vevs &Symmetry& Properties\\
         &&of vacuum & \\\hline
   \texttt{EWs}&(0,0,0)&$\Z2\times\Z2^\prime$ &EW Symmetry\\\hline
  \texttt{2-Inert}&$(0,0,v_3)$&$\Z2\times\Z2^\prime$ &SM + 2 DM candidates\\\hline
    \texttt{DM1}&$(0,v_2,v_3)$&$\Z2$ &2HDM + 1 DM candidates\\\hline
\texttt{DM2}&$(v_1,0,v_3)$&$\Z2^\prime$ &2HDM + 1 DM candidates\\\hline
\texttt{F0DM1}&$(0,v_2,0)$&$\Z2$ &1 DM candidates + massless fermions
    \\\hline
\texttt{F0DM2}&$(v_1,0,0)$&$\Z2^\prime$ &1 DM candidates + massless fermions
    \\\hline
\texttt{F0DM0}&$(v_1,v_2,0)$&None &No DM candidate + massless fermions
    \\\hline
\texttt{N}&$(v_1,v_2,v_3)$&None &3HDM no DM candidate
    \\\hline
\texttt{sCPv}&$(v_1e^{i \xi_1},v_2e^{i \xi_2},v_3)$&None &Spontaneous
CP violation\\\hline
\multicolumn{4}{c}{ }\\\hline
\texttt{F0DM0'}&$(v_1,i v_2,0)$&None &No DM candidate + massless fermions
\\\hline
  \end{tabular}
  \caption{Possible neutral vacua. The top of the table has all vacua found in
    Ref.~\cite{Hernandez-Sanchez:2020aop}; the last line corresponds to a new vacuum. (See text for explanation.)}
  \label{tab:1}
\end{table}

We want to get the conditions where the \texttt{2-Inert} minimum lies below
all other neutral minima.
Ref.~\cite{Hernandez-Sanchez:2020aop} shows that \texttt{DM1} and \texttt{DM2}
are always above \texttt{2-Inert}.
As will be explained below, we found that guaranteeing that the \texttt{2-Inert} minimum lies below
the other minima on the first part of Table~\ref{tab:1} does \textit{not} guarantee that
it lies below our new minimum.
This is easy to see with the following argument.
The difference between the new \texttt{F0DM0'} case and the \texttt{F0MD0} case in
Ref.~\cite{Hernandez-Sanchez:2020aop} is that the former can be obtained from
the latter with the substitution $\phi_2 \rightarrow i \phi_2$.
As can be seen from Eqs.~\eqref{Z2Z2quadratic}-\eqref{Z2Z2quartic},
this corresponds to $\lambda_{10}'' \rightarrow - \lambda_{10}''$
and
$\lambda_{12}'' \rightarrow - \lambda_{12}''$.

Now, in the \texttt{F0MD0} case, ensuring that \texttt{2-Inert} lies below involves
$\lambda_7 + 2 \lambda_{10}''$.
But, since the vev $(v_1, i v_ 2, 0)$ is allowed, and since it is obtainable
through $\lambda_{10}'' \rightarrow - \lambda_{10}''$, we must also
study $\lambda_7 - 2 \lambda_{10}''$.
This plausibility argument will be fully proved both analytically
and numerically below.

\subsection{Charge breaking vacua}
\label{subsec:CB-vacua}

In addition to normal vacua, where the photon is massless,
there are also charge breaking (CB) vacua,
and one must also guarantee stability against then.
Those discussed in Ref.~\cite{Hernandez-Sanchez:2020aop} can be found
in the upper part of Table~\ref{tab:2}.
\begin{table}[htb]
  \centering
  \begin{tabular}{|c|c|}\hline
    Name &vevs \\ \hline
   \texttt{CB1}&
$
\left( \begin{array}{c}
u_1\\
c_1
\end{array} \right)
\ \ 
\left( \begin{array}{c}
u_2\\
c_2
\end{array} \right)
\ \ 
\left( \begin{array}{c}
0\\
c_3
\end{array} \right)
$
 \\ \hline
  \texttt{CB2}&
$
\left( \begin{array}{c}
u_1\\
0
\end{array} \right)
\ \ 
\left( \begin{array}{c}
u_2\\
c_2
\end{array} \right)
\ \ 
\left( \begin{array}{c}
0\\
c_3
\end{array} \right)
$
 \\\hline
    \texttt{CB3}&
$
\left( \begin{array}{c}
u_1\\
c_1
\end{array} \right)
\ \ 
\left( \begin{array}{c}
u_2\\
0
\end{array} \right)
\ \ 
\left( \begin{array}{c}
0\\
c_3
\end{array} \right)
$
\\\hline
\texttt{CB4}&
$
\left( \begin{array}{c}
u_1\\
c_1
\end{array} \right)
\ \ 
\left( \begin{array}{c}
u_2\\
c_2
\end{array} \right)
\ \ 
\left( \begin{array}{c}
0\\
0
\end{array} \right)
$
 \\\hline
\texttt{CB5}&
$
\left( \begin{array}{c}
0\\
c_1
\end{array} \right)
\ \ 
\left( \begin{array}{c}
u_2\\
c_2
\end{array} \right)
\ \ 
\left( \begin{array}{c}
0\\
c_3
\end{array} \right)
$
 
    \\\hline
\texttt{CB6}&
$
\left( \begin{array}{c}
u_1\\
c_1
\end{array} \right)
\ \ 
\left( \begin{array}{c}
0\\
c_2
\end{array} \right)
\ \ 
\left( \begin{array}{c}
0\\
c_3
\end{array} \right)
$
 
    \\\hline
\texttt{CB7}&
$
\left( \begin{array}{c}
u_1\\
0
\end{array} \right)
\ \ 
\left( \begin{array}{c}
u_2\\
0
\end{array} \right)
\ \ 
\left( \begin{array}{c}
0\\
c_3
\end{array} \right)
$

    \\\hline
\texttt{CB8}&
$
\left( \begin{array}{c}
u_1\\
0
\end{array} \right)
\ \ 
\left( \begin{array}{c}
0\\
0
\end{array} \right)
\ \ 
\left( \begin{array}{c}
0\\
c_3
\end{array} \right)
$

    \\\hline
\texttt{CB9}&
$
\left( \begin{array}{c}
0\\
0
\end{array} \right)
\ \ 
\left( \begin{array}{c}
u_2\\
0
\end{array} \right)
\ \ 
\left( \begin{array}{c}
0\\
c_3
\end{array} \right)
$
 \\\hline
\multicolumn{2}{c}{ }\\\hline
\texttt{F0CB}&
$
\left( \begin{array}{c}
u_1\\
c_1
\end{array} \right)
\ \ 
\left( \begin{array}{c}
u_2\\
-\frac{u_1^* u_2}{c_1^*}
\end{array} \right)
\ \ 
\left( \begin{array}{c}
0\\
0
\end{array} \right)
$

\\\hline
  \end{tabular}
  \caption{Possible charge breaking (CB) vacua.
The top of the table has all vacua found in Ref.~\cite{Hernandez-Sanchez:2020aop};
the last line corresponds to a new vacuum. (See text for explanation.)
In all cases, the vacua shown explicitly are assumed to be non-vanishing and unrelated;
except on the last line, where the explicit relation $u_1^* u_2 + c_1^* c_2 = 0$ holds. }
  \label{tab:2}
\end{table}

Ref.~\cite{Hernandez-Sanchez:2020aop} studies the tapdole (stationarity) 
equations for CB vacua in their equations (3.34)-(3.38).
Typically, those equations yield two solutions for the quartic parameters $m_{11}^2$
and/or $m_{22}^2$.
And, forcing their equality gives a constraint on  
the quartic couplings which, when used in the value of the potential at the minimum,
imposes that the CB vacua 
\texttt{CB1}-\texttt{CB9} always lie above the \texttt{2-Inert} vacuum \cite{Hernandez-Sanchez:2020aop}.

As we will show below, we have found both analytically and numerically
that there is, however, a further CB vacuum,
which we dub \texttt{F0CB},
corresponding to the last line
of Table~\ref{tab:2}.
In hindsight,
this arises because, under those very specific conditions
($c_3=0$ and $u_1^* u_2 + c_1^* c_2 = 0$),
both $m_{11}^2$ and $m_{22}^2$ are unequivocally determined from the tadpole
equations, with no further constraint on the quartic parameters.
Again, the existence of this new and independent vacuum will be proved
numerically and analytically below.

\subsection{Numerical  minimization}
\label{subsec:numerical}

The most general vacuum to be compared with the  \texttt{2-Inert} minimum
may be parametrized
as~\cite{Faro:2019vcd}
\begin{equation}
  \label{eq:1}
  \phi_1=\sqrt{r_1}
  \begin{pmatrix}
    \sin\alpha_1\\
    \cos\alpha_1\ e^{i \beta_1}
  \end{pmatrix},\quad
  \phi_2= \sqrt{r_2}
  e^{i \gamma}
  \begin{pmatrix}
    \sin\alpha_2\\
    \cos\alpha_2\ e^{i \beta_2}
  \end{pmatrix},\quad
  \phi_3=\sqrt{r_3}
  \begin{pmatrix}
    0\\
    1
  \end{pmatrix} .
\end{equation}
The main result of Ref.~\cite{Hernandez-Sanchez:2020aop}
is that they only worry about the \texttt{F0DM1}, \texttt{F0DM2} and
\texttt{F0DM0} cases to make sure that the \texttt{2-Inert} is the
global minimum.  They give conditions on the parameters of the
potential, but for \texttt{F0DM0} it is simpler to make a numerical
comparison of the value of the potential in the two situations.

To cross check their results we used the following method.
We start by choosing some point in the $(m^2_{ii}, \lambda_k)$
parameter space of Eqs.~\eqref{Z2Z2quadratic}-\eqref{Z2Z2quartic}.
We ensure that the point satisfies the BFB conditions explained in
Section~\ref{subsec:theorcond}.
Then, we impose
\begin{equation}
  \label{eq:3}
  V_{\texttt{2Inert}} < V_{\texttt{X}}\, ,
\end{equation}
for all the $\texttt{X}$ mentioned in Ref.~\cite{Hernandez-Sanchez:2020aop}.
Next we use the multi-step procedure explained in
Ref.~\cite{Bree:2023ojl}, utilizing CERN's Minuit
library~\cite{james:1975dr} in order to minimize the potential.
In this method,
we minimize the potential starting from a large number of random
initial conditions for the parameters in Eq.~(\ref{eq:1}).
We found that out of 10000 points satisfying BFB and Eq.~\eqref{eq:3},
there were still 161 (1.6\%) that had a lower minima than
$V_{\texttt{2Inert}}$.
Note that this is not a numerical precision problem,
because for those cases where $V_{\texttt{2Inert}}$ was indeed the global
minimum we got precisely the value of the potential, and also
$r_1=r_2=0$ in the notation of Eq.~(\ref{eq:1}).
After adding the constraint
\begin{equation}
  \label{eq:8}
  V_{\texttt{2Inert}} < V_{\texttt{F0DM0'}}\, ,
\end{equation}
we still find points whose minimum lies below \texttt{2Inert}.
But, after adding the constraint
\begin{equation}
  \label{eq:8888}
  V_{\texttt{2Inert}} < V_{\texttt{F0CB}}\, ,
\end{equation}
our extensive minimization procedure no longer finds any
global minimum below $V_{\texttt{2Inert}}$.

It turns out that all the new points correspond to $r_3=0$ and,
thus, we study next that case analytically in detail.

\section{Solving the minimization equations when \texorpdfstring{$r_3=0$}{}}
\label{sec:mineq}

Here we perform an analytical study of the minimization,
which agrees with the numerical results discussed previously.

\subsection{The potential for \texorpdfstring{$r_3=0$}{}}

For both new cases we have found numerically,
\texttt{F0DM0'} and \texttt{F0CB},
one has $r_3=0$.
So,
to have a better understanding of the situation,
we consider the parameterization of Eq.~(\ref{eq:1}) with $r_3=0$.
Certainly, there are redundant angles,
as we shall see in a moment.
With these conditions the potential reads,
\begin{equation}
V
= V_1
+\frac{1}{4}\lambda_7 r_1 r_2 f_7(\alpha_{+},\alpha_{-},\beta)
+ \frac{1}{2}\lambda''_{10} r_1 r_2
f_{10}(\alpha_{+},\alpha_{-},\beta,\gamma),
  \label{eq:10}
\end{equation}
where
\begin{equation}
V_1 = m^2_{11} r_1 + m^2_{22} r_2+ \lambda_1 r_1^2   + \lambda_2
r_2^2+ \lambda_4 r_1 r_2\, ,
\end{equation}
and we have defined
\begin{equation}
  \label{eq:11}
  \alpha_{+}=\alpha_1+\alpha_2,\quad
  \alpha_{-}=\alpha_1-\alpha_2,\quad
  \beta=\beta_1-\beta_2,
\end{equation}
already indicating that we need less angles to describe this
situation, and where we also have defined,
\begin{align}
  \label{eq:12}
  f_7(\alpha_{+},\alpha_{-},\beta)=&
  2 - \cos(2 \alpha_{+}) (-1 + \cos\beta) + \cos(2 \alpha_{-}) (1 + \cos\beta),
\nonumber\\[+2mm]
f_{10}(\alpha_{+},\alpha_{-},\beta,\gamma)=&
\left[\cos(2 \alpha_{+}) (-1 + \cos\beta) + 2 \cos\beta
+ \cos(2 \alpha_{-}) (1 + \cos\beta)\right]
\cos(\beta - 2 \gamma)\nonumber\\
&
- 4 \cos(\alpha_{-}) \cos(\alpha_{+}) \sin\beta \sin(\beta - 2 \gamma).
 \end{align}

\begin{figure}[htpb!]
\centering
\includegraphics[width = 0.4\textwidth]{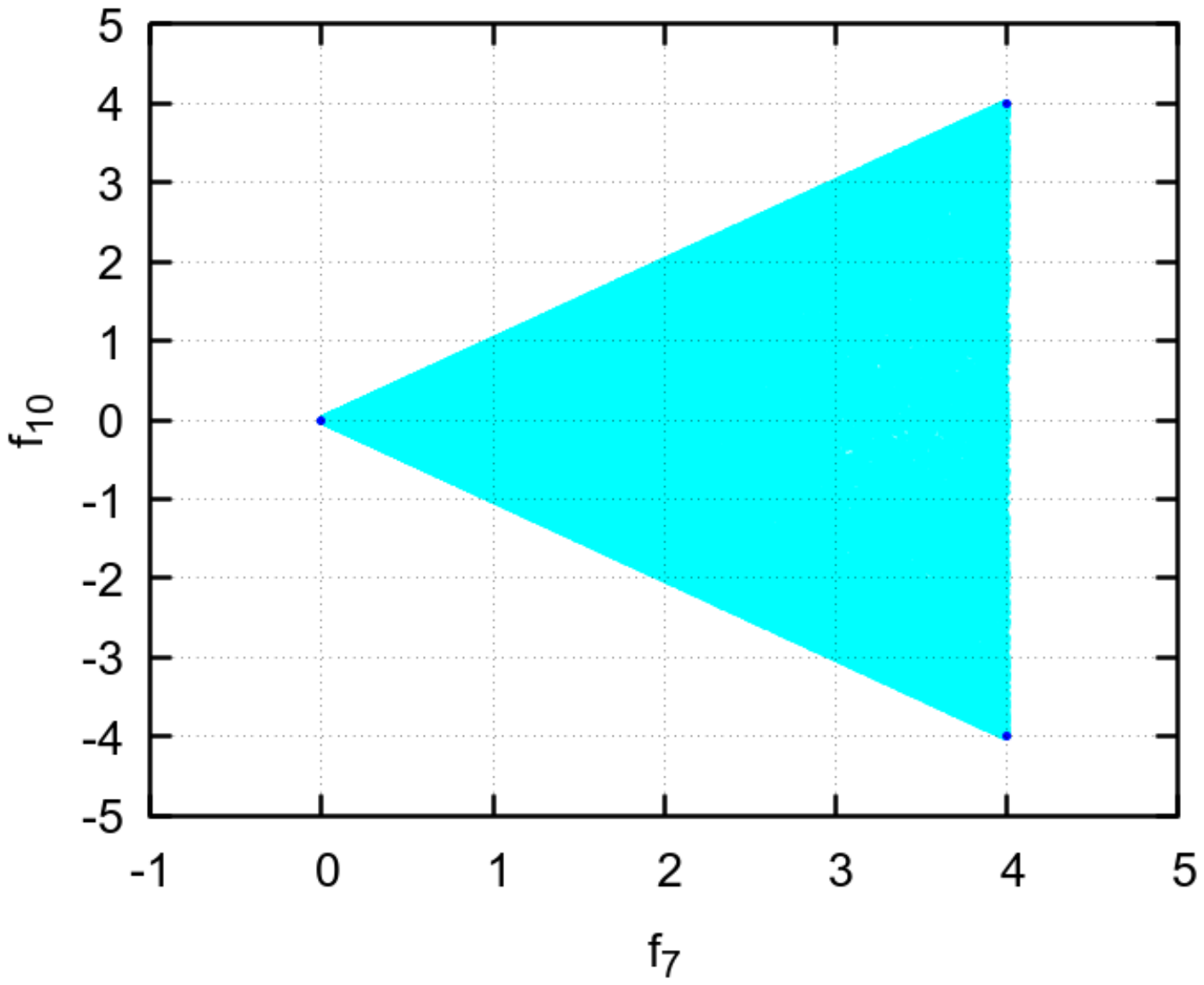}
\caption{Possible values of the functions $f_7$ and $f_{10}$ required for the minimization of the angular part.\label{fig:f7Vsf10}}
\end{figure}

In Fig.~\ref{fig:f7Vsf10}, we plot $f_7$ versus $f_{10}$, for random values of
$\alpha_+$, $\alpha_-$, $\beta$, and $\gamma$.
We see that $x=f_7 \in (0, 4)$,
while $y=f_{10} \in (-4,4)$, lying between the lines
$y = - x$ and $y=+x$.
The angular part of our potential is of the form
$g(x,y) = a x + b y$, with $a= \lambda_7/4$ and $b= \lambda_{10}''/2$.
It is easy to show that $g(x,y)$ cannot have extrema in the interior of the triangle
in Fig.~\ref{fig:f7Vsf10}.
And, assuming $\lambda_7 \neq \pm 2 \lambda_{10}''$,
it must have its extrema at the vertices of the triangle.
The possibilities are, thus,
\begin{eqnarray}
(f_7, f_{10}) = (0,0)
&\ \ \Longrightarrow \ \ &
g(0,0)=0\, ,
\label{0,0}
\\
(f_7, f_{10}) = (4,4)
&\ \ \Longrightarrow \ \ &
g(4,4) = \lambda_7 + 2 \lambda_{10}''\, ,
\label{4,4}
\\
(f_7, f_{10}) = (4,-4)
&\ \ \Longrightarrow \ \ &
g(4,-4) = \lambda_7 - 2 \lambda_{10}''\, .
\label{4,-4}
\end{eqnarray}
The true minimum depends on which of Eqs.~\eqref{0,0}-\eqref{4,-4} lies the lowest.
Notice that this is the only part that depends on $\lambda_7$ and/or $\lambda_{10}''$;
the candidates for extrema do \textit{not} depend on $\lambda_7$ and/or $\lambda_{10}''$,
for they can only lie at the vertices of the triangle regardless.

We conclude that
\begin{eqnarray}
\lambda_7 > 2 |\lambda_{10}''| \hspace{7ex}
&\ \ \Longrightarrow \ \ &
V_\textrm{min} = V_1\, ,
\label{case:0}
\\
\lambda_7 < 2 |\lambda_{10}''| \ \ \ \textrm{and}\ \ \ 
\lambda_{10}'' < 0
&\ \ \Longrightarrow \ \ &
V_\textrm{min} = V_1 + r_1 r_2 (\lambda_7 + 2 \lambda_{10}'')\, ,
\label{case:+}
\\
\lambda_7 < 2 |\lambda_{10}''| \ \ \ \textrm{and}\ \ \ 
\lambda_{10}'' > 0
&\ \ \Longrightarrow \ \ &
V_\textrm{min} = V_1 + r_1 r_2 (\lambda_7 - 2 \lambda_{10}'')\, .
\label{case:-}
\end{eqnarray}
Equations~\eqref{case:0}-\eqref{case:-} correspond, respectively,
to the values obtained for the potential:
i) in our new charge breaking case \texttt{F0CB};
ii) in the case \texttt{F0DM0} of  Ref.~\cite{Hernandez-Sanchez:2020aop};
and
iii) in our new case \texttt{F0DM0'}.

By looking at the extrema conditions for $f_7$ and $f_{10}$ in
Appendix~\ref{app:f7f10_cond}, one can show that
\eqref{case:0} occurs for one of the following angle combinations:
\begin{eqnarray}
\textit{i)}
&\ \ &
\beta =0 \ \ \textrm{and}\ \  \cos{\alpha_-}=0\, ,
\\
\textit{ii)}
&\ \ &
\beta =\pi \ \ \textrm{and}\ \   \cos{\alpha_+}=0\, ,
\\
\textit{iii)}
&\ \ &
\sin\beta \neq 0 \ \ \textrm{and}\ \ \cos(2 \alpha_+) = \cos(2 \alpha_-) = -1\, .
\end{eqnarray}

Similarly,
by looking at the extrema conditions for $f_7$ and $f_{10}$, one can show that
\eqref{case:+} and \eqref{case:-} occur, respectively, for one of the following angle combinations:
\begin{eqnarray}
\textit{i)}
&\ \ &
\beta =0 \ \ \textrm{and}\ \  \sin{\alpha_-}=0 \ \ \textrm{and}\ \  \cos(2 \gamma) = \pm 1\, ,
\\
\textit{ii)}
&\ \ &
\beta =\pi \ \ \textrm{and}\ \  \sin{\alpha_+}=0 \ \ \textrm{and}\ \  \cos(2 \gamma) = \pm 1\, ,
\\
\textit{iii)}
&\ \ &
\sin\beta \neq 0 \ \ \textrm{and}\ \ (\alpha_+,\alpha_-) = (0,0), (\pi,\pi)
\ \ \textrm{and}\ \ \cos(2 \beta - 2 \gamma) = \pm 1\, ,
\\
\textit{iv)}
&\ \ &
\sin\beta \neq 0 \ \ \textrm{and}\ \ (\alpha_+,\alpha_-) = (0,\pi), (\pi,0)
\ \ \textrm{and}\ \ \cos(2 \gamma) = \pm 1\, .
\end{eqnarray}

We have crossed checked our results with our numerical method of
finding the global minimum and they completely agree.

\section{Setting up the scan}
\label{sec:scan}

In the previous sections we discussed in detail how to ensure that
our \texttt{2-Inert} is indeed the global minimum.
We used implicitly the relations found for this minimum.
Namely, 
we take $v_1=v_2=0$, $v_3=v$ and find
\begin{equation}
  \label{eq:2}
  m_h^2=2 \lambda_3 v^2, \quad v^2=-\frac{m_{33}^2}{\lambda_3}\, ,
\end{equation}
requiring $\lambda_3 >0$ (already needed for BFB) and $m_{33}^2 <0$.
The fields can be parametrized as
\begin{equation}
  \label{eq:fields}
  \phi_1=
  \begin{pmatrix}
    H_1^+\\
    \tfrac{1}{\sqrt{2}} \left( H_ 1 + i A_ 1\right)
  \end{pmatrix},\quad
  \phi_2= 
 \begin{pmatrix}
    H_2^+\\
    \tfrac{1}{\sqrt{2}} \left( H_ 2 + i A_ 2\right)
  \end{pmatrix},\quad
  \phi_3=
  \begin{pmatrix}
    G^+\\
    \tfrac{1}{\sqrt{2}} \left( v + h + i G_0 \right)
  \end{pmatrix} .
\end{equation}
Since the vacuum does not break the 
$\Z2\times\Z2$ symmetry
in Eq.~\eqref{eq:20},
all states are unmixed; they are already in the mass basis.
Moreover, $G^0$ and $G^+$ are the would-be Goldstone bosons,
which, in the unitary gauge, become the longitudinal components
of the $Z^0$ and $W^+$ gauge bosons, respectively.

It proves useful\footnote{Our definition of
$\Lambda_1$ agrees with the vertex in Fig.~9b of
\cite{Hernandez-Sanchez:2020aop}, but not with their Eqs.~(2.39),
(5.7)-(5.8).}
to define \cite{Hernandez-Sanchez:2020aop}
\begin{eqnarray}
\Lambda_1 =
\tfrac{1}{2} \left( \lambda_4 + \lambda_7 + 2 \lambda''_{10} \right)\, ,
&\hspace{4ex}&
\bar\Lambda_1 =
\tfrac{1}{2} \left( \lambda_4 + \lambda_7 - 2 \lambda''_{10} \right)\, ,
\\
\Lambda_2 =
\tfrac{1}{2} \left( \lambda_6 + \lambda_9 + 2 \lambda''_{12} \right)\, ,
&\hspace{4ex}&
\bar\Lambda_2 =
\tfrac{1}{2} \left( \lambda_6 + \lambda_9 - 2 \lambda''_{12} \right)\, ,
\\
\Lambda_3 =
\tfrac{1}{2} \left( \lambda_5 + \lambda_8 + 2 \lambda''_{11} \right)\, ,
&\hspace{4ex}&
\bar\Lambda_3 =
\tfrac{1}{2} \left( \lambda_5 + \lambda_8 - 2 \lambda''_{11} \right)\, .
\end{eqnarray}
The other masses are given by
\begin{align}
  m_{H_{1}}^2=&\ m_{11}^2 + \frac{1}{2}\left(\lambda_5 + \lambda_8 + 2
    \lambda''_{11} \right) v^2 \equiv m_{11}^2 + \Lambda_3 v^2\, ,\\ 
  m_{A_{1}}^2=&\ m_{11}^2 + \frac{1}{2}\left(\lambda_5 + \lambda_8 - 2
    \lambda''_{11} \right)v^2 \equiv m_{11}^2 + \bar{\Lambda}_3 v^2\, ,\\ 
  m_{H^\pm_{1}}^2=&\ m_{11}^2 + \frac{1}{2} \lambda_5 v^2\, ,\\[+2mm]
  m_{H_{2}}^2=&\ m_{22}^2 + \frac{1}{2}\left(\lambda_6 + \lambda_9 + 2
    \lambda''_{12} \right) v^2 \equiv m_{22}^2 + \Lambda_2 v^2\, ,\\ 
  m_{A_{2}}^2=&\ m_{22}^2 + \frac{1}{2}\left(\lambda_6 + \lambda_9 - 2
    \lambda''_{12} \right)v^2 \equiv m_{22}^2 + \bar{\Lambda}_2 v^2\, ,\\ 
  m_{H^\pm_{2}}^2=&\ m_{22}^2 + \frac{1}{2} \lambda_6 v^2\, .
\label{m_Hpm_2}
\end{align}
The conditions for a local minimum are
\begin{align}
  v^2=-\frac{m_{33}^2}{\lambda_3}>0\, ,
\quad
\Lambda_2 > -m_{22}^2/v^2\, ,
\quad
\Lambda_3 > - m_{11}^2/v^2\, ,
\end{align}
but, if we take the masses as input parameters, these will be
automatically satisfied. The value of the potential at the minimum is
\begin{align}
  V_{\texttt{2Inert}} = - \frac{m_{33}^4}{4\lambda_3}\, .
\label{V_2Inert}
\end{align}
As mentioned, we ensure that $ V_{\texttt{2Inert}}$ lies below
the value of the potential at the other local extrema,
whose explicit expressions
can be found in Appendix~\ref{app:masses}.

We take
\begin{equation}
v = 246\,\text{GeV}\, , \ \ \  m_h = 125\,\text{GeV}\, ,
\end{equation}
as fixed inputs.
We follow Ref.~\cite{Hernandez-Sanchez:2020aop} and choose as our free parameters
\begin{equation}
m_{H_1}^2, m_{H_2}^2, m_{A_1}^2,m_{A_2}^2,
m_{H_1^\pm}^2, m_{H_2^\pm}^2,
\Lambda_1, \Lambda_2, \Lambda_3,
\lambda_1, \lambda_2, \lambda_4, \lambda_7\, .
\label{input_param}
\end{equation}
All other parameters of the scalar potential can be extracted from
Eq.~\eqref{eq:2}-\eqref{m_Hpm_2}.
We choose random values for the remaining parameters in the
set of Eq.~\eqref{input_param},
in the ranges
\begin{align}
&\Lambda_1,\, \Lambda_2,\, \Lambda_3,\, \lambda_1,\, \lambda_2,\, \lambda_4,\, \lambda_7  \in \pm\left[10^{-3},10\right];
\nonumber\\[8pt]
&m_{H_1},\, m_{H_2},\, m_{A_1},\,m_{A_2}\,
\in \left[50,1000\right]\,\text{GeV};
\nonumber\\[8pt]
&
m_{H_1^\pm},\,m_{H_2^\pm}\,
\in \left[70,1000\right]\,\text{GeV},
\label{eq:scanparameters}
\end{align}
with the chosen condition that $m_{H_1}<m_{H_2}$, without loss of generality.
The lower limit on the mass of the charged scalars comes from Ref.~\cite{Pierce:2007ut}.
Although this bound has not been established within the context of the
current model of two component DM, we take it as a conservative
lower bound on the masses of all charged scalars.

For the interactions with fermions and gauge bosons,
it is assumed that all such SM fields transform into themselves under
$\Z2\times\Z2$.
Thus,
Eq.~\eqref{eq:20} implies that \textit{all} fermion fields only couple with $\phi_3$.
This is a so-called Type-I model.
Since the Yukawa couplings are identical to the SM ones,
there are no FCNCs at tree-level. Moreover,
as the charged scalars are inert and lack coupling to fermions,
they bypass many of the constraints found in the usual 2HDMs.
Notably, flavour bounds on the charged scalar masses, such as from $B \to X_s \gamma$ are trivially satisfied. As the fields in Eq.~\eqref{eq:fields} are already in the mass basis,
there are thus three sectors: the dark-$\Z2$ sector,
constituted by the fields in $\phi_1$;
the dark-$\Z2^\prime$ sector,
constituted by the fields in $\phi_2$;
and the active or SM sector,
constituted by the fields in $\phi_3$
and all SM fermions.
Connections among different sectors can only occur due to gauge bosons or due
to the cubic and quartic interactions of the Higgs potential.

We have generated \texttt{FeynMaster} \cite{Fontes:2019wqh,Fontes:2021iue},
and, through it, \texttt{FeynRules} \cite{Alloul:2013bka}
model files which yield all Feynman
rules\footnote{The complete and consistent set of all
Feynman Rules for this model may be found at the url
\url{https://porthos.tecnico.ulisboa.pt/~romao/Work/arXiv/3HDMZ2xZ2-Inert2/}.}
and were used to generate
an input file for \texttt{micrOMEGAs 6.0.5} \cite{Alguero:2023zol}.

Our numerical scan proceeds in the following fashion.
We start by taking a random value for the parameters \eqref{input_param} within
the intervals \eqref{eq:scanparameters}. The values of $\Lambda_1,\, \Lambda_2,\, \Lambda_3,\, \lambda_1,\, \lambda_2,\, \lambda_4$,
and $\lambda_7$ where scanned log-uniformly; all other input parameters were scanned uniformly.
For the constraints in this section, all parameters were actually scanned over.
For the plots in section \ref{sec:results}, extensive dedicated scans around points yielding the
correct relic density were performed. In some DM mass regions, points were easier
to generate; in other regions, a good point was very hard to come by. We continued this process
until a reasonable density in each plot was obtained. As mentioned above,
we apply the constraints from BFB and global minimum.
Then, as a sanity check, we confirm numerically that indeed no lower-lying minimum is found. 
Next,
we impose perturbative unitarity \cite{Moretti:2015cwa,Bento:2022vsb},
and compliance with the experimental oblique radiative parameter $STU$
\cite{Baak:2014ora},
utilizing the general formulae in \cite{Grimus:2008nb}.
We calculate all processes at lowest non-trivial order and take all input parameters
at the electroweak scale.

\section{Further experimental constraints}
\label{sec:further}

\subsection{Collider constraints}

We have adapted our in-house scanning program including
already the latest LHC bounds on the $h_{125}$ signal strengths with
the full Run~2 data collected at~13~TeV,
for the different production and decay modes,
following the ATLAS results\footnote{The generic agreement between
ATLAS and CMS measurements implies that our conclusions
will not be altered significantly if we use instead CMS results or a suitable
combination and ATLAS and CMS results.}
summarized in Fig.~3 of
\cite{ATLAS:2022vkf}.
For comparison with these collider experiments, we consider only the decay
contributions of the lowest non-vanishing order in perturbation theory
when comparing with the coupling modifiers of the most recent ATLAS fit
result Ref.~\cite{ATLAS:2022vkf}, within $2\sigma$.
For a specific production mechanism and decay channel,
the Higgs signal strength is defined as:
\begin{equation}
\mu_i^f =
\left(\frac{\sigma_i^{\text{3HDM}}(pp\to h) }{\sigma_i^{\text{SM}}(pp\to h)}\right)
\left(\frac{\text{BR}^{\text{3HDM}}(h\to f)}{\text{BR}^{\text{SM}}(h\to f)}\right)\, ,
	\label{e:ss}
\end{equation}
with the subscript `$i$' corresponding to the production mode
and the superscript `$f$' to the decay channel of the $125\, \text{GeV}$ Higgs scalar.
The relevant production
mechanisms considered are gluon fusion~($ggF$),
vector boson fusion~(VBF),
associated production with a vector boson ($VH$, $V = W$ or $Z$),
and
associated production with a pair of top quarks ($ttH$).
The SM cross
section for the gluon fusion process is calculated using HIGLU
\cite{Spira:1995mt}, and for the other production mechanisms we use
the prescription of Ref.~\cite{LHCHiggsCrossSectionWorkingGroup:2016ypw}. The final states in the decay channels considered are $f=\,W\,W,\, Z\,Z,\,
b\,\overline{b}, \gamma\,\gamma$ and $\tau^+\tau^-$.
The upper limit on the Higgs total decay width is set
by Ref.~\cite{CMS:2019ekd} at:
\begin{equation}
    \Gamma_\textrm{tot}\leq 9.1\, \text{MeV}\, .
\end{equation}

We forbid decays of SM gauge bosons into the new scalars by enforcing:
\begin{equation}
m_{H_i}+m_{H_i^\pm} \geq m_W^\pm\, , \hspace{2ex}
m_{A_i}+m_{H_i^\pm} \geq m_W^\pm\, , \hspace{2ex}
m_{H_i}+m_{A_i} \geq m_Z\, , \hspace{2ex}
2 m_{H_i^\pm}\geq m_Z.
\end{equation}
Taking into account the LEP 2 results re-interpreted for the
I(1+1)HDM, we exclude the region of masses where the following
conditions are simultaneously satisfied
\cite{Lundstrom:2008ai} $(i=1,2)$:
\begin{equation}
m_{A_i} \leq 100\, \text{GeV},\quad
m_{H_i} \leq 80\, \text{GeV},\quad
|m_{A_i}-m_{H_i}| \geq 8 \, \text{GeV}.
\end{equation}
In order to evade the bounds from long-lived charged particle searches given in Ref.~\cite{Heisig:2018kfq}, we set
the upper limit on the charged scalar lifetime of $\tau \leq 10^{-7}\text{s}$.

The surviving points were then passed through
\texttt{HiggsTools} 1.1.3 \cite{Bahl:2022igd},
imposing current bounds from searches for additional
scalars.

\subsection{DM constraints}

Our results for the relic density, scattering amplitudes and
annihilation cross section are obtained using the implementation
of this model in \texttt{micrOMEGAs 6.0.5} \cite{Alguero:2023zol},
which we have constructed.

We calculate the dark matter relic density as the sum of the contributions
of each DM candidate:
\begin{equation}
    \Omega_T h^2 = \Omega_1 h^2 + \Omega_2 h^2\,,
\end{equation}
and apply the limits obtained by the Planck experiment
\cite{Planck:2018vyg}\footnote{Alternatively,
one could adopt a more permissive range. Indeed, some models have been studied
where loop effects can induce corrections to the relic density which are
of the order of 10\% \cite{Banerjee:2019luv,Banerjee:2021hal}; this lead the authors of
\cite{Belanger:2021lwd} to consider instead the augmented range
$0.094 < \Omega_T h^2 < 0.142$. We will keep to the
Planck constraint in \eqref{Planck}.}:
\begin{equation}
    \Omega_{T} h^2= 0.1200\pm 0.0012\, .
\label{Planck}
\end{equation}

The strongest direct detection limit from dark matter-nucleon
scattering is currently provided by Ref.~\cite{LZ:2022lsv}.
To compare directly with the experimental limit, we follow the
method presented in  \cite{Belanger:2014bga} of computing the
normalized cross section of DM on a point-like nucleus (taken to be xenon)
\begin{equation}
\sigma_{\text{SI}}^{\text{Xe},k}=
\frac{4 \mu_k^2}{\pi}\frac{\left(Z f_p + (A-Z)f_n\right)^2}{A^2},
\end{equation}
with $\mu_k$ the reduced mass of the DM candidate and
$f_p$, $f_n$ the amplitudes for protons and neutrons.
As there are two dark matter candidates, we rescale
the obtained cross section for each DM candidate by the
relative density of the component:
\begin{equation}
\sigma_{\text{SI}}^{\text{r},k} =
\sigma_{\text{SI}}^{\text{Xe},k}\ \xi_k\, , 
\end{equation}
where
\begin{equation}
\xi_k = \frac{\Omega_k}{\Omega_T}\, .
\end{equation}

To deal with indirect detection constraints, we follow closely
the strategy adopted in \cite{Belanger:2021lwd}.
We start by using our \texttt{micrOMEGAs 6.0.5} model implementation
in order to calculate the thermally averaged cross section for DM annihilation
(or co-annihilation, or DM conversion)
times velocity $\langle \sigma v \rangle$.
Current Fermi-LAT limits for indirect DM detection through photons \cite{Fermi-LAT:2015att}
range from $\langle\sigma v\rangle\approx 3\times 10^{-26} \text{cm}^3/\text{s}$,
for light DM,
to around $\langle\sigma v\rangle\approx  10^{-25} \text{cm}^3/\text{s}$,
for heavier DM.
As in \cite{Belanger:2021lwd}, we find that the $\langle \sigma v \rangle$
which are in reach of Fermi-LAT arise mainly from
annihilation decays into $VV$.
So we sum only the  $WW$ and $ZZ$ final states,
assuming a similar spectrum, which we dub $\langle\sigma v\rangle_{VV}$,
and we limit it to the 95\% CL bound from Fermi-LAT \cite{Fermi-LAT:2015att}.

Searches for anti-protons with AMS-02 \cite{AMS:2016oqu,AMS:2016brs} yield the most
stringent constraints for WIMP DM.
Again, we consider $\langle\sigma v\rangle_{VV}$ and use the bounds obtained in
\cite{Reinert:2017aga}.
We also comment briefly on limits from H.E.S.S. \cite{HESS:2022ygk},
a Cherenkov gamma ray, ground-based telescope,
which points to the central region of the Milky Way. 
This yields the strongest ID constraints for DM masses
from $\sim 500\,\textrm{GeV}$ upwards.

\section{Some interesting processes}
\label{sec:processes}

\begin{figure}[htbp!]
\centering
\includegraphics[width=0.75\textwidth]{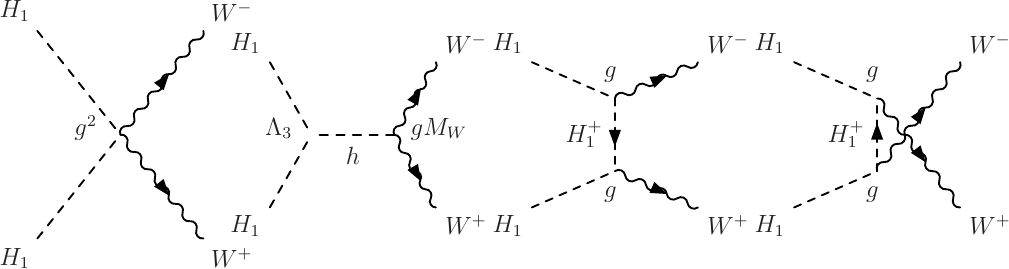}
\caption{Feynman diagrams for $H_1 H_1 \rightarrow WW$.
}
\label{H1H1TOWW}
\end{figure}
%
%
Before we proceed, it is interesting to describe some of the classes of processes achievable
in this very rich model.
Indeed, we have:
\begin{itemize}
\item (vanilla) annihilation: e.g. $H_1 H_1 \rightarrow b \bar{b}$,
\item co-annihilation: e.g. $A_1 H_1 \rightarrow b \bar{b}$,
\item co-scattering: e.g. $H_1^+ W^- \rightarrow A_1 Z$,
\item DM conversion: e.g. $H_2 H_2 \rightarrow H_1 H_1$.
\end{itemize}
There are, however, no semi-annihilations processes $x_i x_j \rightarrow x_k \textrm{SM}$,
such as appears in \cite{DEramo:2010keq}.
Fig.~\ref{H1H1TOWW}
shows the Feynman diagrams for $H_1 H_1 \rightarrow WW$.
The Feynman diagrams for $H_1 H_1 \rightarrow ZZ$ are obtained
from Fig.~\ref{H1H1TOWW} with the substitutions $W^\pm \rightarrow Z$,
$H_1^\pm \rightarrow A_1$.

\begin{figure}[htbp!]
\centering
\includegraphics[width=0.40\textwidth]{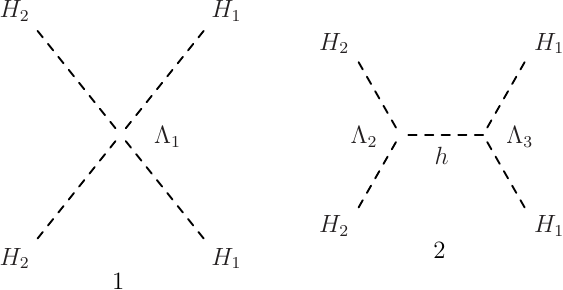}
\caption{Feynman diagrams for the DM conversion processes
$H_2 H_2 \rightarrow H_1 H_1$.
}
\label{22TO11}
\end{figure}

\begin{figure}[htbp!]
\centering
\includegraphics[width=0.4\textwidth]{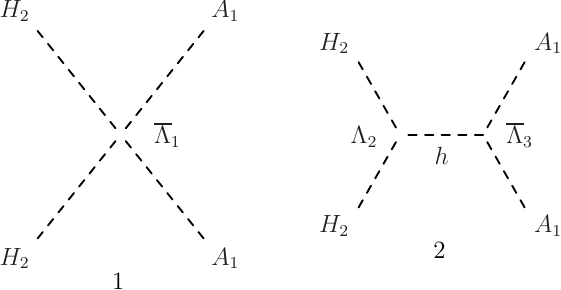}
\caption{Feynman diagrams for the DM conversion processes
$H_2 H_2 \rightarrow A_1 A_1$.
}
\label{22TOA1A1}
\end{figure}

\begin{figure}[htbp!]
\centering
\includegraphics[width=0.40\textwidth]{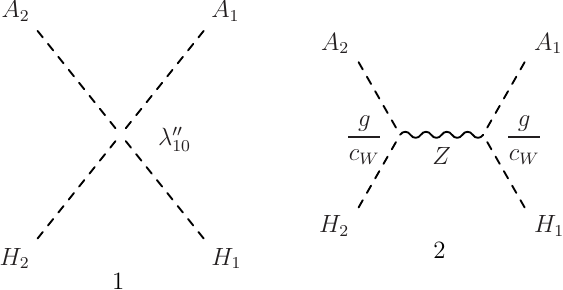}
\caption{Feynman diagrams for the DM conversion processes
$H_2 A_2 \rightarrow H_1 A_1$.
}
\label{2A2TO1A1}
\end{figure}

This model has DM conversion processes between the DM sector 1 and the
DM sector 2:
\begin{itemize}
\item $H_2 H_2 \rightarrow H_1 H_1$: quartic ($\sim \Lambda_1$), and
through $h$ ($\sim \Lambda_2 \Lambda_3$);
\item $H_2 H_2 \rightarrow A_1 A_1$: quartic ($\sim \bar{\Lambda}_1$), and
through $h$ $(\sim \Lambda_2 \bar{\Lambda}_3)$;
\item  $H_2 A_2 \rightarrow H_1 A_1$: quartic ($\sim \lambda''_{10}$),
and through $Z$ ($\sim g^2$).
\end{itemize}
We show the Feynman diagrams for these DM conversion processes
in Figs.~\ref{22TO11}, \ref{22TOA1A1}, and \ref{2A2TO1A1}.

The numerical results to be discussed below include all processes, which arise out of
our implementation of the model in \texttt{micrOMEGAs 6.0.5}.
The authors of Ref.~\cite{Hernandez-Sanchez:2020aop} choose to concentrate on the mass region
$m_h/2 < m_{H_1} < 80 \textrm{GeV}$ and $ m_{H_2} \simeq 100 \textrm{GeV}$.
We extend significantly the analysis by considering all available parameter space.
Not surprisingly, we find different conclusions.
In particular, we find many situations in which both DM components can contribute equally
to the relic density.
We also find wide regions of parameter space where
the possibility that the lighter DM component is mainly probed through direct nuclear recoil
while de heavier DM component is probed in indirect DM detection does not hold.
This will be discussed next.

\section{Results and discussion}
\label{sec:results}

In the IDM, the DM mass is constrained to two regions.
The reason is the following.
The annihilation of the DM particle into $WW$ (equivalent to that in Fig.~\ref{H1H1TOWW})
is controlled by a gauge coupling and, thus, it is not tunable.
For the most part, it leads to a decay rate so high that it depletes the
IDM DM candidate, making its relic density under-abundant.
This is avoided for $m_\textrm{dm} \gtrsim 500 \textrm{GeV}$ if all ``dark'' scalars
($H$, $A$, and $H^\pm$) have similar masses.
Below the $W$ threshold, the annihilation proceeds mostly into
$b \bar{b}$; this depends on the DM-Higgs coupling, which is tunable to comply
with the relic density.
However, that requires large couplings, which are precluded by direct detection.
The exception occurs around $m_\textrm{dm} \simeq m_h/2$ where the annihilation
has a resonance, allowing for a fit to the relic density with a coupling low
enough to comply with direct detection constraints.

\begin{figure}[htbp!]
\centering
\includegraphics[width=0.55\textwidth]{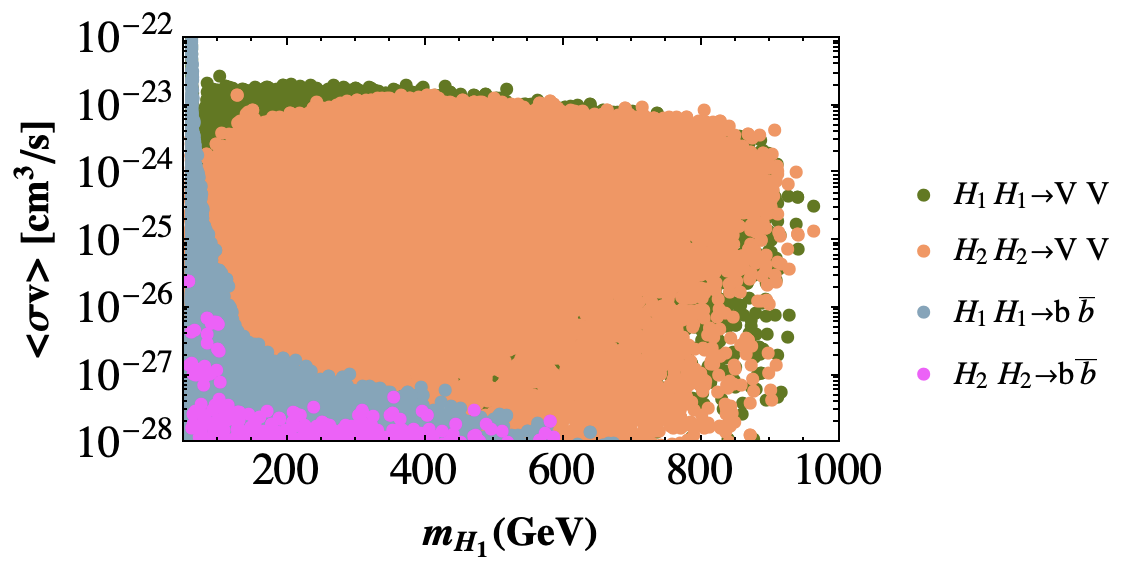}
\caption{$\langle\sigma v\rangle$ for
$H_1$ ($H_2$) annihilation into $b \bar{b}$ and $VV$ as a function of $m_{H_1}$.
The colour codes are in the figure.
}
\label{sigv_H1_and_H2}
\end{figure}

\begin{figure}[htbp!]
\centering
\includegraphics[width=0.45\textwidth]{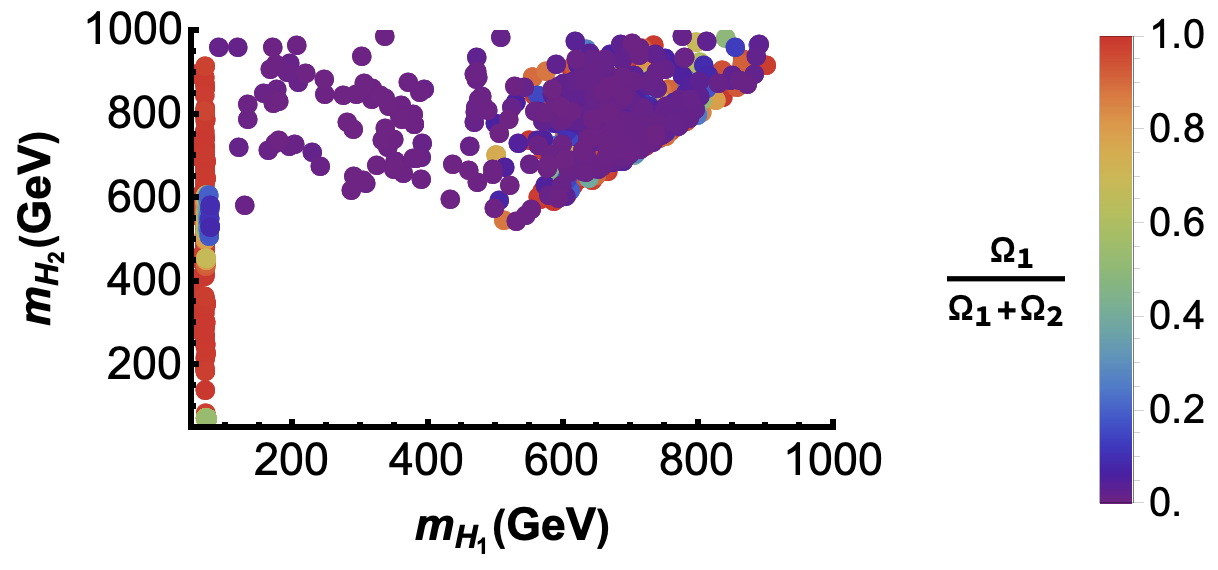}
\hspace{5mm}
\includegraphics[width=0.45\textwidth]{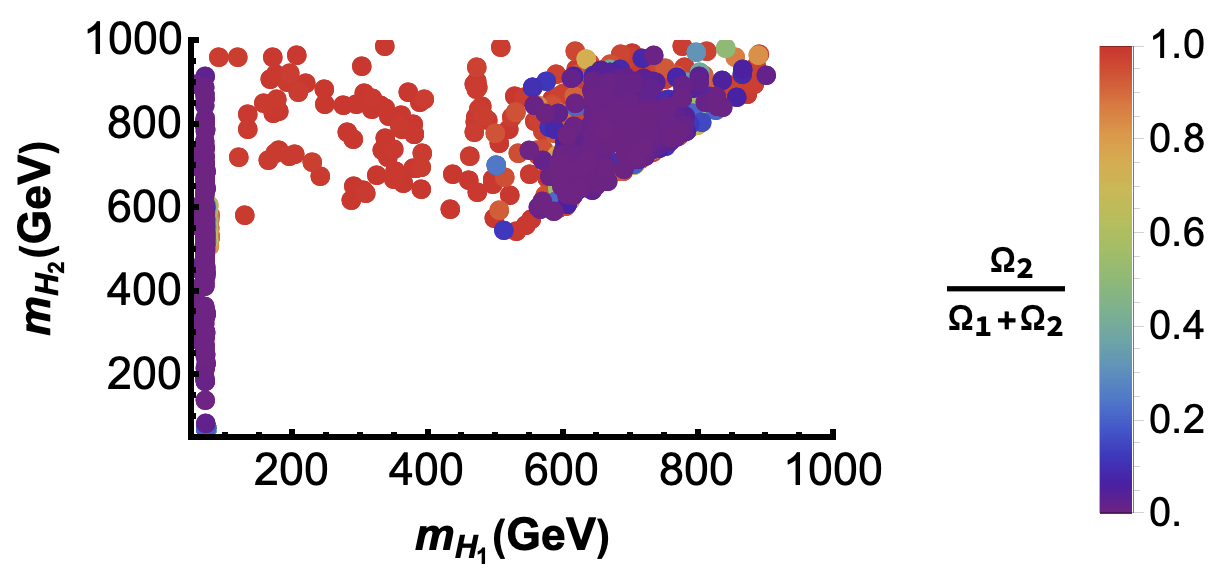}
\caption{Range of allowed ($m_{H_1}, m_{H_2}$) masses with a ``temperature''
colour code for $\Omega_1/\Omega_T$ ($\Omega_2/\Omega_T$) on the left (right).
Both plots have the same information but were included to aid the eye.
These points have passed all theory, collider and astrophysical constraints.
}
\label{mH1mH2_relic}
\end{figure}

The situation is both similar and different in our two component DM
$\Z2\times\Z2$ 3HDM.
This is best seen with the help of Fig.~\ref{sigv_H1_and_H2},
where we show the values of $\langle\sigma v\rangle$ for
$H_1$ ($H_2$) annihilation into $b \bar{b}$ and $VV$ as a function of $m_{H_1}$.
First, we note that above the $W$ threshold
$\langle\sigma v\rangle_{VV} \gg \langle\sigma v\rangle_{bb}$ for both $H_1$ and $H_2$.
Second, for $H_1$, $\langle\sigma v\rangle_{bb}$ can be very large
for low values of $m_{H_1}$, as in the IDM.
Thus, if $H_1$ were the only DM component, combining relic density
and direct detection constraints would lead to the same mass regions
as in the IDM.
However, we can force $m_{H_1}$ into the intermediate mass range, by requiring that
it is $H_2$ which is mostly responsible for the relic density.
This can be seen in Fig.~\ref{mH1mH2_relic}.
Notice also that for small $m_{H_1}$ there are two possibilities.
For any value of $m_{H_2}$,
one can have $H_1$ be the major relic density component.
On the other hand,
for $ 500 \textrm{GeV}  \lesssim m_{H_2} \lesssim  600 \textrm{GeV}$,
one can have $H_2$ be the major relic density component.
Moreover,
one can find quite a number of interesting points
where $\Omega_1 \simeq \Omega_2$.
These occur for the $H_1$ mass regions which would be allowed in the IDM.
The reason is simple; $H_1$ would be able to allow for all the relic density,
and one can tune it down for 50\%, while tuning the $H_2$ parameters to account
for the remainder 50\%.
Finally, we note that if a DM component has a mass between $\sim 100 \textrm{GeV}$ and
$\sim 500 \textrm{GeV}$,
then it is guaranteed to give a very suppressed contribution to the
total relic density.

There is one relevant issue concerning Fig.~\ref{sigv_H1_and_H2}.
One might worry that there are significant contributions from
$\langle\sigma v\rangle_{hh}$.
We have checked that this is the case if we do not impose that $\Omega_T$
must equal the measured relic density.
However, once we impose the Planck limit in Eq.~\eqref{Planck},
the dominant contributions are those shown in Fig.~\ref{sigv_H1_and_H2}.
Albeit in a different model, this is also what was found in Ref.~\cite{Belanger:2021lwd}.
This will be important to the discussion on indirect detection below.

\begin{figure}[htbp!]
\centering
\includegraphics[width=0.37\textwidth]{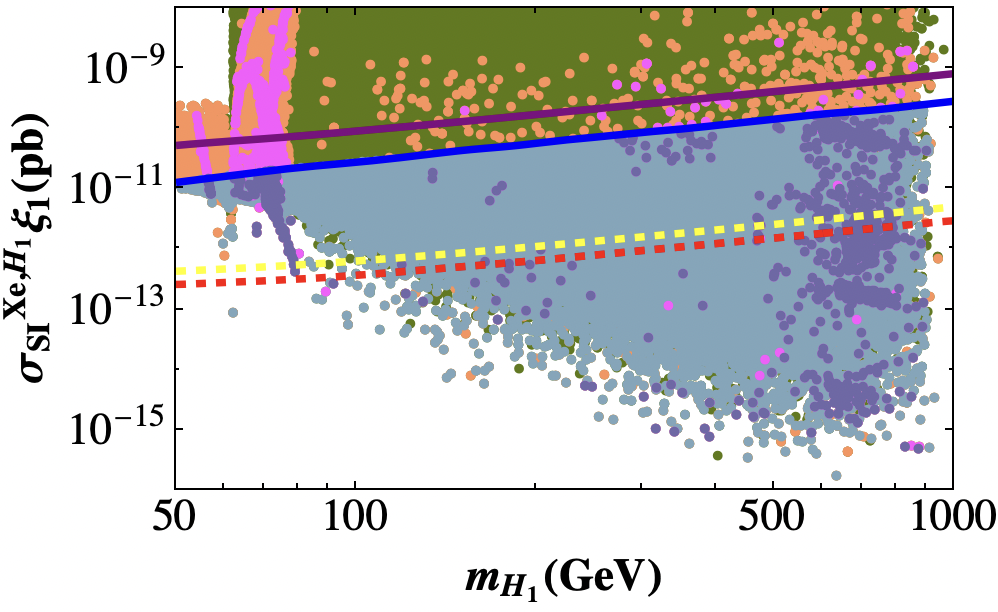}
\hspace{3mm}
\includegraphics[width=0.14\textwidth]{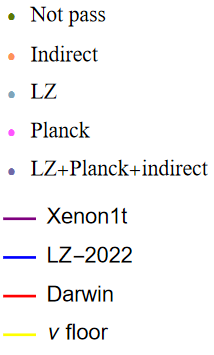}
\hspace{3mm}
\includegraphics[width=0.37\textwidth]{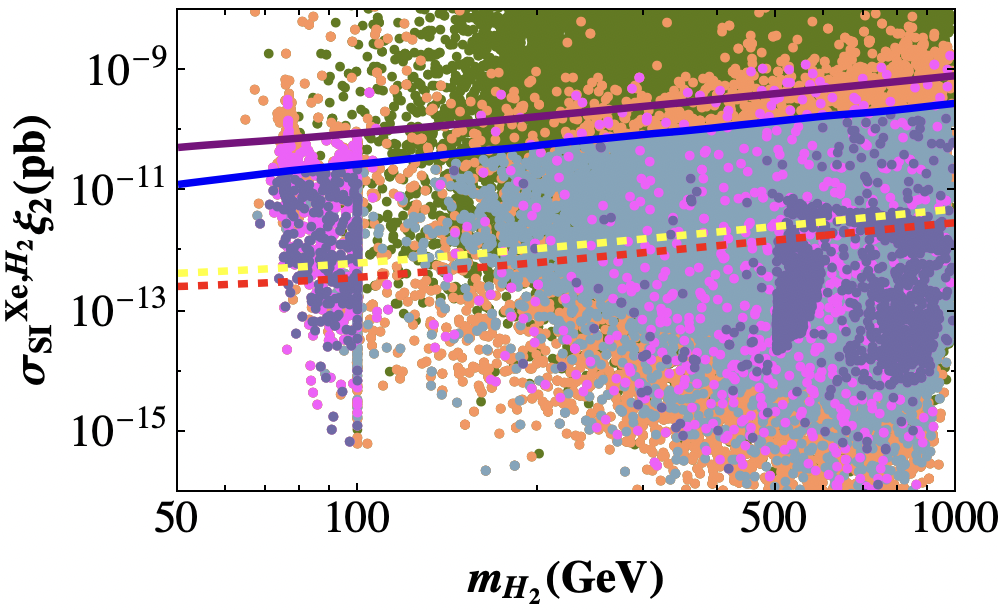}
\caption{Direct detection constraints on $H_1$ ($H_2$) on the left (right) figure.
See text for details.
}
\label{direct}
\end{figure}

But first, we turn to the constraints from direct detection shown in Fig.~\ref{direct}.
The lines shown have the following origins:
i) the solid purple line refers to the XENON1T \cite{XENON:2018voc},
which is included as recast limits inside \texttt{micrOMEGAs 6.0.5};
ii) the solid blue line refers to the current LUX-ZEPLIN (LZ) bound
\cite{LZ:2022lsv};
iii) the dashed red line corresponds to the expected reach
of the DARWIN experiment \cite{DARWIN:2016hyl};
iv) the dashed yellow line corresponds to the neutrino floor,
as presented in \cite{OHare:2021utq}.
As for the points shown,
all have passed theory and collider constraints.
The colour code refers to the additional astrophysical constraints as follows:
i) the green points have passed  theory and collider constraints,
but failed all of the relic density, direct and indirect limits;
ii) the orange points have agreement with Fermi-LAT's indirect detection
bounds \cite{Fermi-LAT:2015att}, but fail both Planck and LZ;
iii) the gray points Pass LZ, but fail Planck (some pass Fermi-LAT;
some do not);
iv) the pink points achieve the correct relic density,
but failed either LZ or Fermi-LAT;
finally, v) the dark-purple points pass all current astrophysical
constraints.

\begin{figure}[htbp!]
\centering
\includegraphics[width=0.4\textwidth]{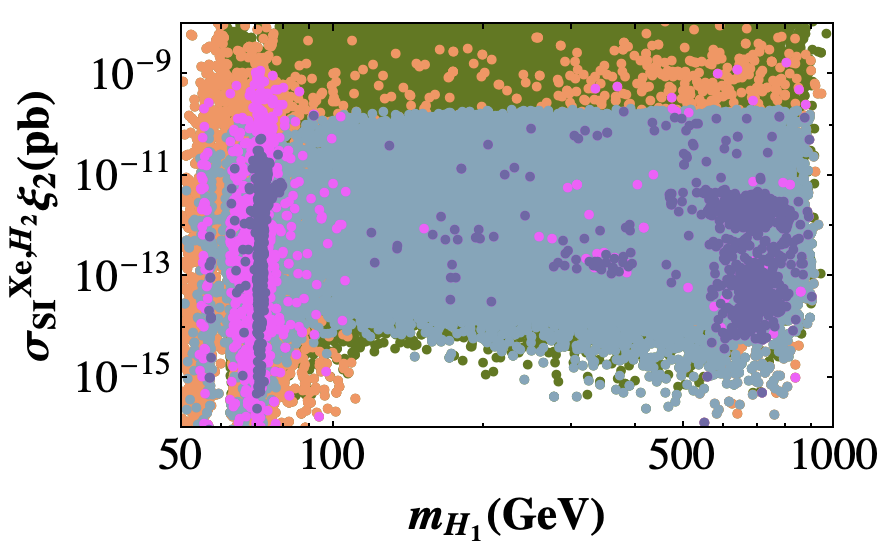}
\caption{Constraints on $\sigma_{\text{SI}}^{\text{Xe}2k}\ \xi_2$  as a function of $m_{H_1}$.
The colours of points have the same meaning as in Fig.~\ref{direct}.
Notice that the curves for experimental constraints are not appropriate
for this graph.
}
\label{direct2mh1}
\end{figure}

Let us concentrate first on Fig.~\ref{direct}-left.
The presence of red and orange points above the LZ line shows that,
in agreement with our previous discussion,
this constraint from direct detection is relevant for low $H_1$ masses.
Remarkably, for such low masses the DARWIN experiment will be able to
exclude many points.
In contrast,
many points with high $H_1$ masses will not be invalidated by DARWIN.
And, since this exclusion is expected to lie below the neutrino floor,
other collider and/or astrophysical probes must be used.
Turning to Fig.~\ref{direct}-right, we see that direct detection
may also constrain $H_2$.
This is more clearly seen in a plot as a function of
$m_{H_1}$, as in Fig.~\ref{direct2mh1}.
However, looking at the pink and dark-purple points in Fig.~\ref{direct2mh1}
for low values of $m_{H_1}$ it is possible that direct
detection probes $H_1$, while it does not affect $H_2$,
in accordance with the special case discussed in
\cite{Hernandez-Sanchez:2020aop}.

Note that the appearance of green points bellow all exclusion lines in one of the
plots in Fig.~\ref{direct} is due to the fact that, although the point passes
the direct detection for the corresponding DM component,
it does not pass it for the other DM component.

We now turn to the constraints arising from indirect detection.
As mentioned, our strategy was to prove that $\langle\sigma v\rangle_{VV}$ dominates for
most of the parameter space, \and use the lines determined by the authors of Ref.~\cite{Reinert:2017aga} together with gamma ray searches for dark matter \cite{Fermi-LAT:2015att,HESS:2022ygk}.
The exception occurs for small masses, where we apply the $\langle\sigma v\rangle_{bb}$ lines also obtained by the authors of Ref.~\cite{Reinert:2017aga} and Fermi-LAT experiment \cite{Fermi-LAT:2015att}.

\begin{figure}[htbp!]
\centering
\includegraphics[width=0.38\textwidth]{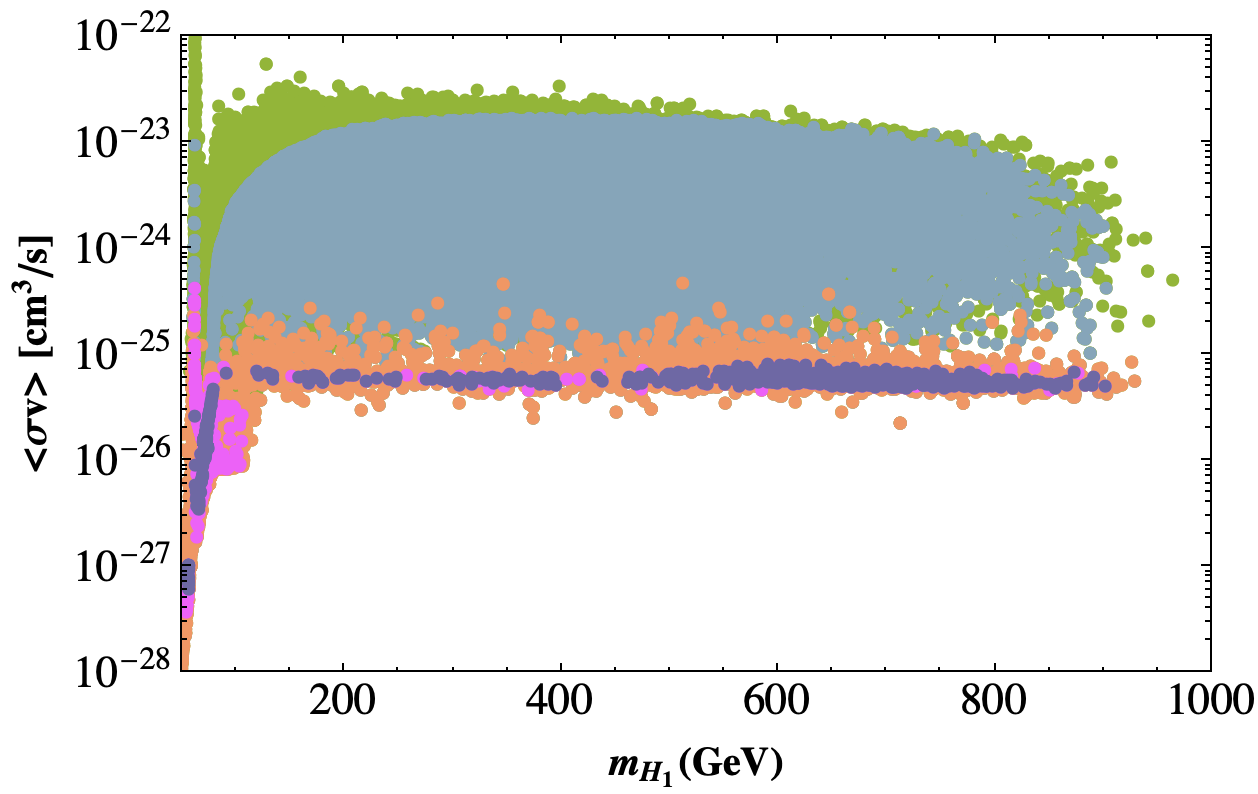}
\hspace{3mm}
\includegraphics[width=0.14\textwidth]{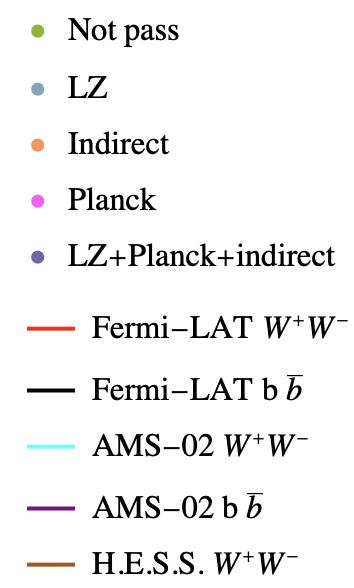}
\hspace{3mm}
\includegraphics[width=0.38\textwidth]{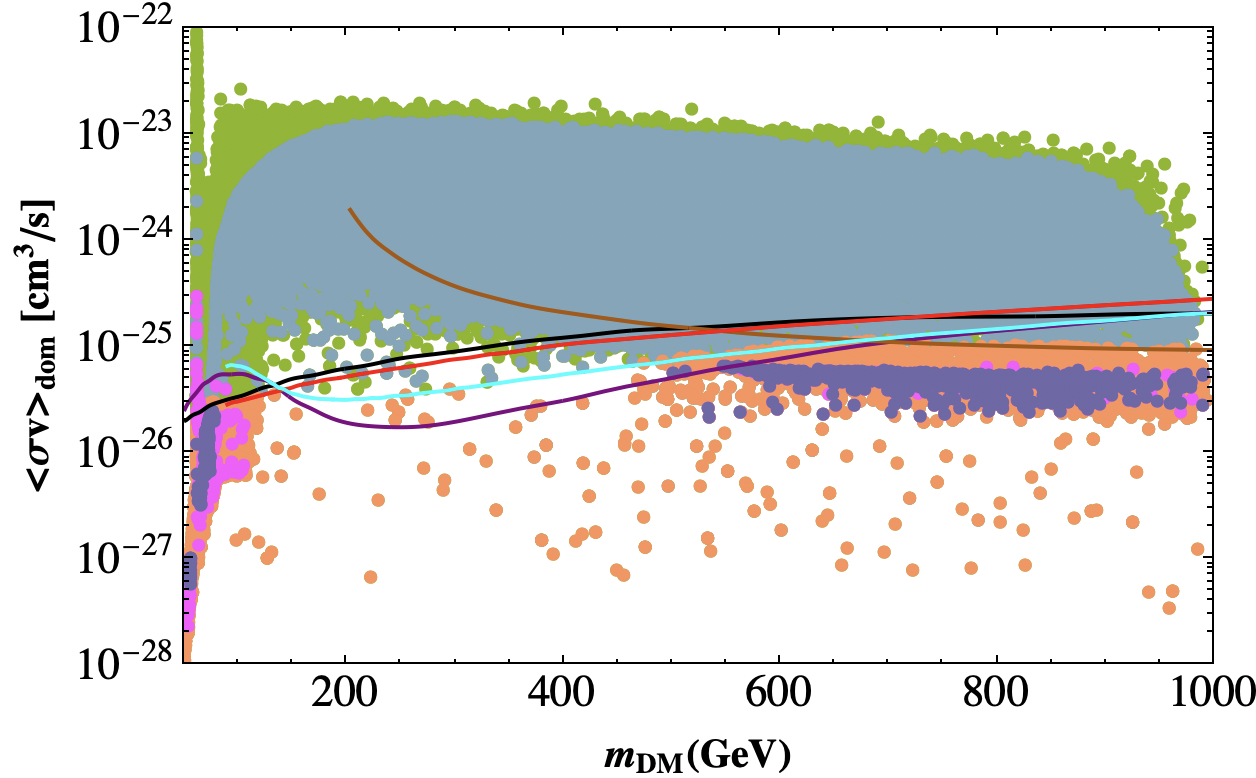}
\caption{The colours of the points have the same meaning as in Fig.~\ref{direct}.
The left figure shows the total
$\langle\sigma v\rangle$ as a function of $m_{H_1}$. 
The right figure shows the dominant contribution to
$\langle\sigma v\rangle$ as a function of the mass of the DM candidate, $m_\textrm{DM}$,
which corresponds to the $\langle\sigma v\rangle$ plotted on the vertical axis.
The lines coming from
Fermi-LAT \cite{Fermi-LAT:2015att} and H.E.S.S. \cite{HESS:2022ygk} assume a Navarro-Frenk-White (NFW)
DM density profile and the AMS-02 \cite{AMS:2016oqu} lines correspond to the conservative approach derived in Ref.~\cite{Reinert:2017aga},
with the colour codes also shown in the figure.
}
\label{indirect1}
\end{figure}

Fig.~\ref{indirect1}-left shows the total $\langle\sigma v\rangle$.
The plots have 200000 points (in fact, we generated 1 million points, but the conclusions
are not altered).
The colours of the points have the same meaning as in Fig.~\ref{direct}.
The red points have passed Planck bounds but may, or not, have passed
the bounds from indirect detection.
However, we found that, out of 2761 points that pass Planck, only 36 are ruled out by
indirect detection, and that this occurs only for masses of $H_1$ between
$\sim 62 \textrm{GeV}$ and $\sim 66 \textrm{GeV}$, where the dominant contribution is
$\langle\sigma v\rangle_{bb}$.
This is also seen on Fig.~\ref{indirect1}-right,
where we plot the dominant contribution to
$\langle\sigma v\rangle$ as a function of the mass of the DM candidate, $m_\textrm{DM}$,
which corresponds to the (dominant) $\langle\sigma v\rangle$ plotted on the vertical axis.
This is to be compared with the exclusion
lines for $\langle\sigma v\rangle_{VV}$ and $\langle\sigma v\rangle_{bb}$
corresponding to $H_1$ or $H_2$; whichever yields the dominant  $\langle\sigma v\rangle$.
The lines come from Ref.~\cite{Reinert:2017aga} and refer to exclusions extrapolated from
Fermi-LAT \cite{Fermi-LAT:2015att} (in red for $VV$ and in black for $bb$),
and from AMS-02 \cite{AMS:2016oqu} (in light-blue for $VV$ and in purple for $bb$).
We also include (in a solid brown line) a limit coming from
H.E.S.S. \cite{HESS:2022ygk},
which is important for DM masses above $\sim 500\,\textrm{GeV}$.
As a result, we can conclude that, except for those very specific 36 points,
the Planck constraints (almost) guarantee that the indirect detection will be
ineffectual.
Again, the black points pass every constraint.

\begin{figure}[htbp!]
\centering
\includegraphics[width=0.45\textwidth]{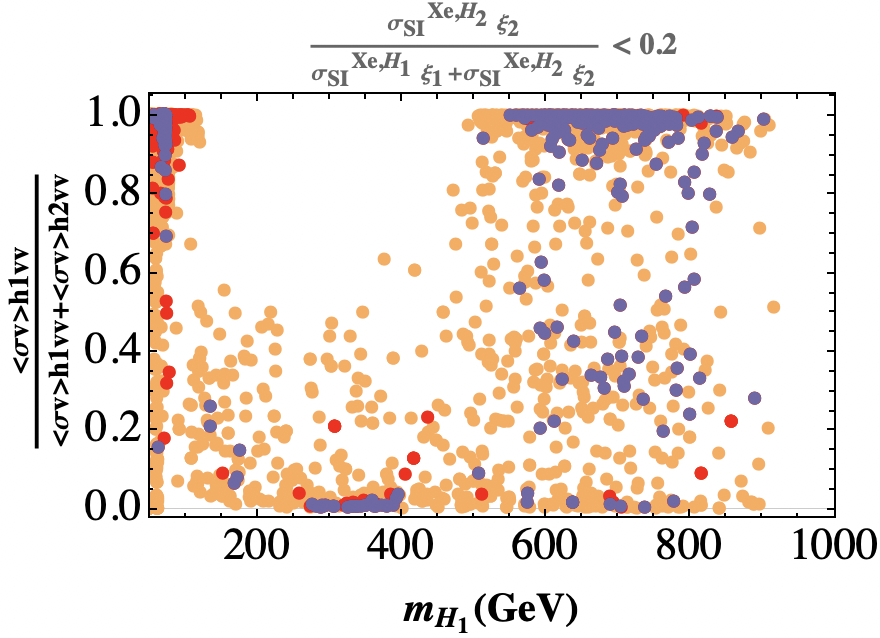}
\hspace{5mm}
\includegraphics[width=0.45\textwidth]{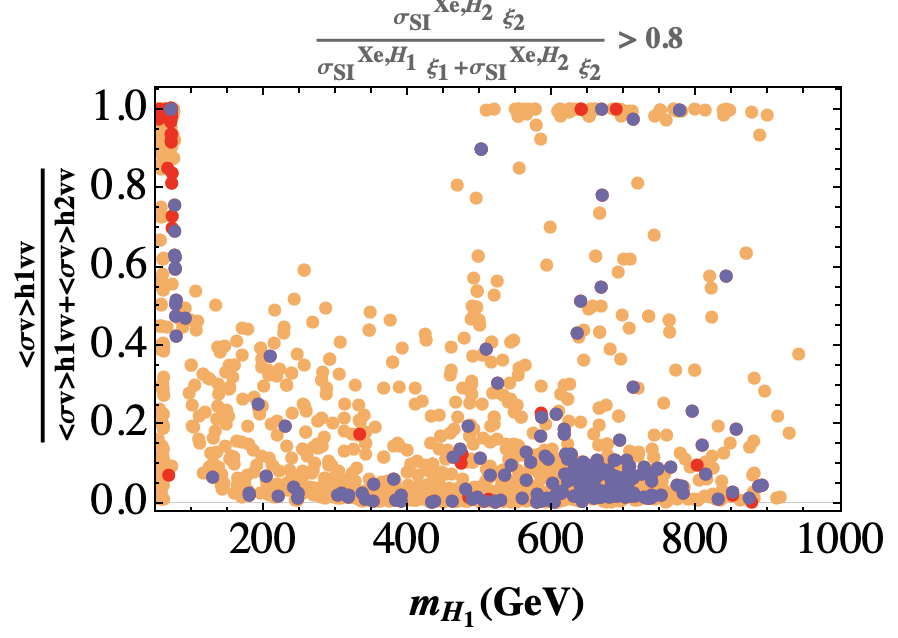}
\caption{The orange points pass the indirect detection bounds; the red ones
achieve the correct density; and the dark-purple points pass all astrophysical constraints.
On the left (right) we show points where direct detection is dominated by $H_1$ ($H_2$).}
\label{indirect2}
\end{figure}

We now turn to the interplay between direct and indirect detection.
Fig.~\ref{indirect2}-left (-right) contains points for which the dominant
direct detection cross-section is due to $H_1$ ($H_2$).
In both, we plot  $\langle\sigma v\rangle_{VV}$ for $H_1$ over the sum of
$\langle\sigma v\rangle_{VV}$ for $H_1$ and $H_2$.
Concentrating on Fig.~\ref{indirect2}-left,
we learn that, whilst direct detection is dominated by $H_1$,
$H_1$ can, in some cases, give the dominant contribution to
indirect detection, while, in other cases, it is $H_2$ which gives
the dominant contribution to indirect detection.
Conversely,
on Fig.~\ref{indirect2}-right,
we learn that, whilst direct detection is dominated by $H_2$,
$H_1$ can, in some cases, give the dominant contribution to
indirect detection, while, in other cases, it is $H_2$ which gives
the dominant contribution to indirect detection.
That is, depending on the parameters of the model,
including $m_{H_1}$, we can have all four possible combinations.

Notice the following feature on both plots in Fig.~\ref{indirect2}.
When $m_{H_1}$ lies roughly between $100 \textrm{GeV}$ and $500 \textrm{GeV}$,
it can never be the dominant contribution to indirect detection signals.
This is also the region where $H_1$ cannot be the dominant contribution to
the relic density. This is the previously referred relation between
relic density and indirect detection.

\begin{figure}[htbp!]
\centering
\includegraphics[width=0.4\textwidth]{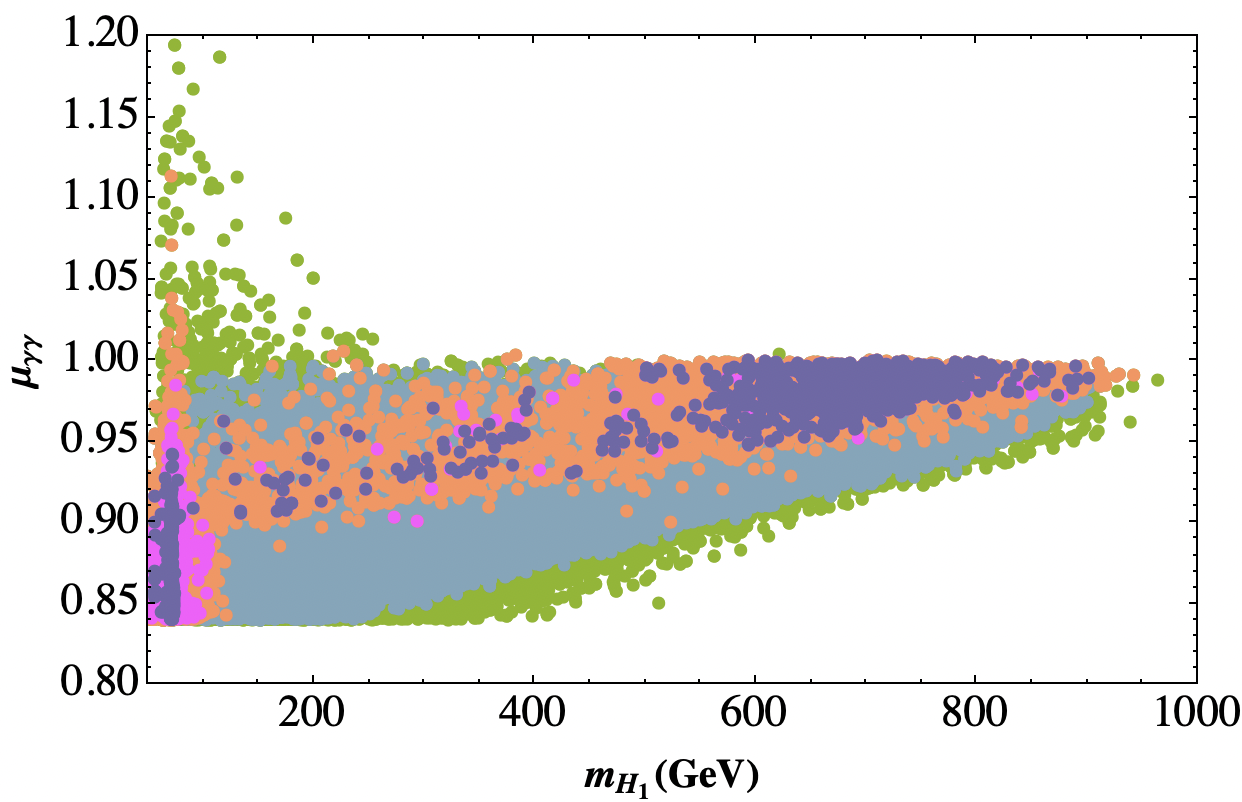}
\caption{Constraints on $\mu_{\gamma\gamma}$ as a function of $m_{H_1}$.
The colours of points have the same meaning as in Fig.~\ref{direct}.
}
\label{hgg}
\end{figure}

Recall,
that since we have only one active scalar, all tree level $125 \textrm{GeV}$ Higgs
couplings are as in the SM model.
It is only in loop mediated (or loop corrections to tree level) decays
that we are sensitive to the dark sectors.
As an example,
we show results for the $h \rightarrow \gamma\gamma$ decay in Fig.~\ref{hgg}.
We start by noticing that, after theoretical and LHC constraints,
there is a region where $\mu_{\gamma\gamma}$ could exceed unity.
Although not completely apparent from the figure,
this region only starts after $M_{H_1} > m_h/2$, and decreases sharply until around
$m_{H_1} \sim 300 \textrm{GeV}$.
This is exactly the same that one finds in the IDM;
compare, for example, with figure 3-left of \cite{Krawczyk:2015vka};
see also \cite{Krawczyk:2013jta}.
However,
this region is already excluded by the LZ direct detection bound.
Thus, in the $\Z2\times\Z2$ 3HDM, current astrophysical bounds force
$\mu_{\gamma\gamma} \lesssim 1$.

\begin{figure}[htbp!]
\centering
\includegraphics[width=0.4\textwidth]{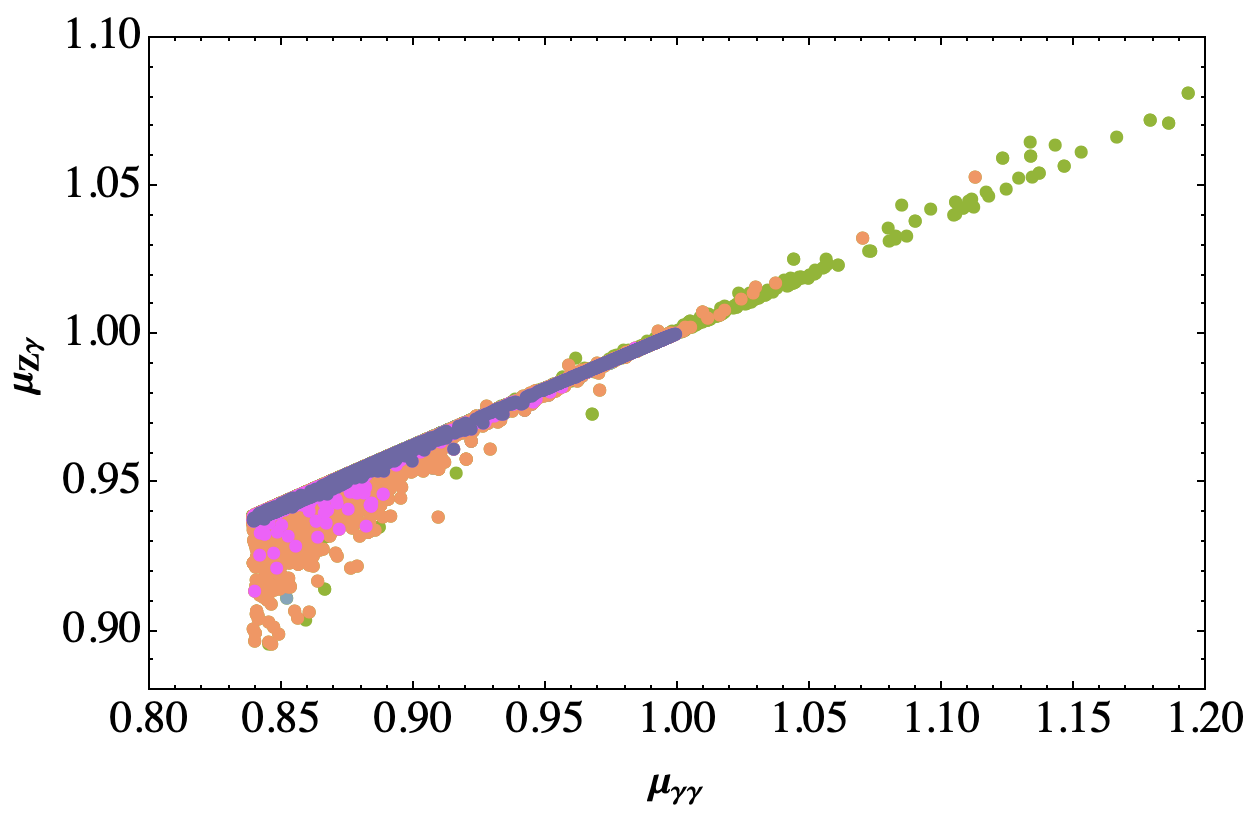}
\caption{Constraints on the $\mu_{\gamma\gamma} - \mu_{Z\gamma}$ plane.
The colours of points have the same meaning as in Fig.~\ref{direct}.
}
\label{hzg}
\end{figure}

As for $h \rightarrow Z \gamma$, 
the current measurement of $ \mu_{Z \gamma} = 2.2 \pm 0.7$
\cite{ATLAS:2020qcv,ATLAS:2023yqk,CMS:2022ahq}, still has large errors.
Nonetheless,
if its central value remained with shrinking errors,
it would constitute a definite sign of new Physics.
We show the results for the $\mu_{\gamma\gamma}$ versus $\mu_{Z \gamma}$
within the $\Z2\times\Z2$ 3HDM in Fig.~\ref{hzg}.
One sees a very strong correlation between
$\mu_{\gamma\gamma}$ and $\mu_{Z\gamma}$,
due to the fact that very similar new virtual charge Higgs diagrams
are involved in both cases.\footnote{It is viable to uncorrelate
$\mu_{Z\gamma}$ from $\mu_{\gamma\gamma}$ in multi-Higgs models
where the $Z$ couples to two different charged Higgses \cite{Boto:2023bpg}.
}
As a result, should a more precise measurement of
$\mu_{Z \gamma}$ uncover new physics,
the $\Z2\times\Z2$ 3HDM would be ruled out together with the SM.

\section{Conclusion}
\label{sec:concl}

Recently there has been renewed interest in multi-component DM models.
We focus our attention on a  $\Z2\times\Z2$ symmetric 3HDM with a double
inert vacuum.
We start by reassessing the possible solutions of the stationarity equations,
making sure that ours is the absolute minimum.
We found new relevant minima, which we dub \texttt{F0DM0}'
and \texttt{F0CB}.
To be certain that we keep all the points that are global minima, we
have to compare not only with the \texttt{F0DM1}, \texttt{F0DM2} and
\texttt{F0DM0}, but also with the new cases, \texttt{F0DM0}'
(Eq.~(\ref{eq:5}) and 
\texttt{F0CB} (Eq.~(\ref{eq:9})).
We obtained explicit expressions for all the cases and
therefore it is easy to compare.
We also include unitarity, BFB and conformance with the oblique parameters
$S$, $T$, and $U$.

After this step, we subject the parameter space of our model to all
current collider constraints, including limits in the 125GeV couplings,
searches for extra scalars and 
flavour observables.
We then concentrate on the implications from relic density, DD and ID
of DM. By performing a wide scan, we found that simple implications obtained
when concentrating on small regions of parameter space cease to be valid,
and a much richer pallet of possibilities emerges.
In particular, we found regions where one can have two DM candidates contributing
equally to the relic density. The whole mass range for a given component can be populated in the $\Z2\times\Z2$ model, even for intermediate mass regions that require the other component to dominate the relic density calculation. We include the future sensitivity of DD experiments that are expected to reach the high mass section of the neutrino fog without being able to invalidate the model, thus requiring complementary probes.

We hope that this work will entice the community to look closer at this stimulating model.
In particular, it will be interesting to explore its implication to more detailed collider
observables, such as monojets with large missing transverse energy, mono-Z or multi-lepton
signals. In addition, one may also seek exploratory predictions for the reach of proposed
future colliders. We leave this for a future publication.

\section*{Acknowledgments}
This work is supported in part
by the Portuguese Funda\c{c}\~{a}o
para a Ci\^{e}ncia e Tecnologia\/ (FCT) under Contracts
CERN/FIS-PAR/0002/2021, UIDB/00777/2020, and UIDP/00777/2020\,;
these projects are partially funded through POCTI (FEDER),
COMPETE, QREN, and the EU.  The work of R. Boto is also supported
	by FCT with the PhD grant PRT/BD/152268/2021.


\appendix

\section{Mass formulas and conditions for local minima}
\label{app:masses}

In this Appendix we present the formulas for the scalar masses under the
various minima of interest. Requiring that the mass squared are positive is akin
to guaranteeing that the corresponding extreme is indeed a local minimum.

\subsection{\texttt{2-Inert}}

This case can be found in Eq.~\eqref{eq:2}-\eqref{V_2Inert} of
Section~\ref{sec:scan}.

%

\subsection{\texttt{F0DM1}}

In this case we have $v_1=0$, $v_2\not=0, v_3=0$. The minimization gives
\begin{align}
  v_2^2=-\frac{m_{22}^2}{\lambda_2} \, ,
\end{align}
implying $m_{22}^2 <0$ as $\lambda_2>0$ from BFB. For the masses we
have
\begin{align}
  m_{H_{1}}^2=& m_{11}^2 + \frac{1}{2}\left(\lambda_4 + \lambda_7 + 2
    \lambda''_{10} \right) v_2^2 \equiv m_{11}^2 + \Lambda_1 v_2^2\, ,\\ 
  m_{A_{1}}^2=& m_{11}^2 + \frac{1}{2}\left(\lambda_4 + \lambda_7 - 2
    \lambda''_{10} \right)v_2^2 \equiv m_{11}^2 + \bar{\Lambda}_1 v_2^2\, ,\\ 
  m_{H^\pm_{1}}^2=& m_{11}^2 + \frac{1}{2} \lambda_4 v_2^2\, ,\\[+2mm]
  m_{H_{2}}^2=& 2 v_2^2 \lambda_2\, ,\\[+2mm]
  m_{H_{3}}^2=& m_{33}^2 + \frac{1}{2}\left(\lambda_6 + \lambda_9 + 2
    \lambda''_{12} \right) v_2^2 \equiv m_{33}^2 + \Lambda_2 v_2^2\, ,\\ 
  m_{A_{3}}^2=& m_{33}^2 + \frac{1}{2}\left(\lambda_6 + \lambda_9 - 2
    \lambda''_{12} \right)v_2^2 \equiv m_{33}^2 + \bar{\Lambda}_2 v_2^2\, ,\\ 
  m_{H^\pm_{3}}^2=& m_{33}^2 + \frac{1}{2} \lambda_6 v_2^2 \, .
\end{align}
We have to require all these masses squared to be positive in order to
have a local minimum. This is easier than finding conditions on the
parameters. The value of the potential at the minimum is
\begin{align}
  V_{\texttt{F0DM1}} = - \frac{m_{22}^4}{4\lambda_2}\, ,
\end{align}
and this has to be compared with $V_{\texttt{2Inert}}$.

\subsection{\texttt{F0DM2}}

In this case we have $v_1\not=0$, $v_2=0, v_3=0$. The minimization gives
\begin{align}
  v_1^2=-\frac{m_{11}^2}{\lambda_1}\, ,
\end{align}
implying $m_{11}^2 <0$ as $\lambda_1>0$ from BFB. For the masses we
have
\begin{align}
  m_{H_{1}}^2=& 2 v_1^2 \lambda_1\, ,\\[+2mm]
  m_{H_{2}}^2=& m_{22}^2 + \frac{1}{2}\left(\lambda_4 + \lambda_7 + 2
    \lambda''_{10} \right) v_1^2 \equiv m_{22}^2 + \Lambda_1 v_1^2\, ,\\ 
  m_{A_{2}}^2=& m_{22}^2 + \frac{1}{2}\left(\lambda_4 + \lambda_7 - 2
    \lambda''_{10} \right)v_1^2 \equiv m_{22}^2 + \bar{\Lambda}_1 v_1^2\, ,\\ 
  m_{H^\pm_{2}}^2=& m_{22}^2 + \frac{1}{2} \lambda_4 v_1^2\, ,\\[+2mm]
  m_{H_{3}}^2=& m_{33}^2 + \frac{1}{2}\left(\lambda_5 + \lambda_8 + 2
    \lambda''_{11} \right) v_1^2 \equiv m_{33}^2 + \Lambda_3 v_1^2\, ,\\ 
  m_{A_{3}}^2=& m_{33}^2 + \frac{1}{2}\left(\lambda_5 + \lambda_8 - 2
    \lambda''_{11} \right)v_1^2 \equiv m_{33}^2 + \bar{\Lambda}_3 v_1^2\, ,\\ 
  m_{H^\pm_{3}}^2=& m_{33}^2 + \frac{1}{2} \lambda_5 v_1^2\, .
\end{align}
We have to require all these masses squared to be positive in order to
have a local minimum. This is easier than finding conditions on the
parameters.
The value of the potential at the minimum is
\begin{align}
  V_{\texttt{F0DM2}} = - \frac{m_{11}^4}{4\lambda_1}.
\end{align}
and this has to be compared with $V_{\texttt{2Inert}}$.

\subsection{\texttt{F0DM0}}

In this case we have $v_1\not=0$, $v_2\not=0, v_3=0$. The minimization gives
\begin{align}
  v_1^2=\frac{\lambda_2 m_{11}^2-\Lambda_1
    m_{22}^2}{\Lambda_1^2-\lambda_1 \lambda_2},\
  v_2^2=\frac{\lambda_1 m_{22}^2-\Lambda_1
    m_{11}^2}{\Lambda_1^2-\lambda_1 \lambda_2}\, ,
\end{align}
requiring $v_{1}^2,v_2^2 >0$. For the masses we
have
\begin{align}
  m_{H_{1}}^2=&\lambda_1 v_1^2+\lambda_2 v_2^2
  -\sqrt{4 \Lambda_1^2 v_1^2 v_2^2 + (\lambda_1 v_1^2 - \lambda_2
    v_2^2)^2} \, ,\\[+2mm] 
  m_{H_{2}}^2=&\lambda_1 v_1^2+\lambda_2 v_2^2
  +\sqrt{4 \Lambda_1^2 v_1^2 v_2^2 + (\lambda_1 v_1^2 - \lambda_2 v_2^2)^2}\, ,\\ 
  m_{H_{3}}^2=& m_{33}^2 + \frac{1}{2}\left(\lambda_5 + \lambda_8 + 2
    \lambda''_{11} \right) v_1^2 + \frac{1}{2}\left(\lambda_6 + \lambda_9 + 2
    \lambda''_{12} \right)v_2^2 \, ,\\ 
  m_{A_{1}}^2=&-2 \lambda''_{10}\left( v_1^2 + v_2^2\right)\, ,\\ 
  m_{A_{2}}^2=& m_{33}^2 + \frac{1}{2}\left(\lambda_5 + \lambda_8 - 2
    \lambda''_{11} \right)v_1^2 + \frac{1}{2}\left(\lambda_6 + \lambda_9 - 2
    \lambda''_{12} \right)v_2^2  \, ,\\
  m_{H^\pm_{1}}^2=& - \frac{1}{2} \left(\lambda_7 +2
    \lambda''_{10}\right)(v_1^2+  v_2^2)\, ,\\[+2mm]
  m_{H^\pm_{2}}^2=& m_{33}^2 + \frac{1}{2}
  \left(\lambda_5 v_1^2+\lambda_6 v_2^2\right)\, .
\end{align}
We have to require all these masses squared to be positive in order to
have a local minimum. This is easier than finding conditions on the
parameters.
The value of the potential at the minimum is
\begin{align}
  V_{\texttt{F0DM0}} = \frac{\lambda_1 m_{22}^4+\lambda_2 m_{11}^4-2
    m_{11}^2m_{22}^2 \Lambda_1}{4(\Lambda_1^2-\lambda_1\lambda_2)}\, ,
\end{align}
and this has to be compared with $V_{\texttt{2Inert}}$.

\subsection{\texttt{F0DM0'}}

Let us take
\begin{equation}
  \label{eq:4}
  \sqrt{r_1}=\frac{v_1}{\sqrt{2}},
  \sqrt{r_2}=\frac{v_2}{\sqrt{2}},r_3=0,\quad
  \alpha_1=\alpha_2=\beta_1=\gamma=0,\quad
  \beta_2=\frac{\pi}{2}.
\end{equation}
This is still along the neutral directions, but in comparison with
\texttt{F0DM0} corresponds to making\footnote{Notice
that this case is not contradiction with Eq.(3.32) of
Ref.~\cite{Hernandez-Sanchez:2020aop}.
Although it looks like a particular case of
\texttt{sCPv}, we checked explicitly by calculating the
full $6 \times 6$ mass matrix for the neutral scalars that indeed that matrix
separates into CP even and CP odd blocks, ensuring CP conservation.} $v_2\to i\, v_2$.
The stationary conditions give,
\begin{align}
  v_1^2=\frac{\lambda_2 m_{11}^2-\bar{\Lambda}_1
    m_{22}^2}{\bar{\Lambda}_1^2-\lambda_1 \lambda_2},\
  v_2^2=\frac{\lambda_1 m_{22}^2-\bar{\Lambda}_1
    m_{11}^2}{\bar{\Lambda}_1^2-\lambda_1 \lambda_2},\
\end{align}
and
\begin{equation}
  \label{eq:5}
  V_{\texttt{F0DM0}'} = \frac{\lambda_1 m_{22}^4+\lambda_2 m_{11}^4-2
    m_{11}^2m_{22}^2 \bar{\Lambda}_1}{4(\bar{\Lambda}_1^2-\lambda_1\lambda_2)}\, ,
\end{equation}
where
\begin{equation}
  \label{eq:6}
  \bar{\Lambda}_1= \frac{1}{2}\left( \lambda_4+\lambda_7 -2
    \lambda_{10}'' \right)\, ,
\end{equation}
was defined before. We require that $v_1^2,v_2^2$ are positive and
also check for the positiveness of the masses. This is not absolutely
necessary, because if $ V_{\texttt{F0DM0}'} <  V_{\texttt{2Inert}}$,
even if it is a saddle-point, it indicates that there should be a
minimum below and the point is not a good point. But our statement is
stronger if we also identify the local minima.
It should be stressed that such a minimum should
exist, as our sufficient BFB conditions were checked for all cases in
our numerical simulation.
For the masses we
have
\begin{align}
  m_{H_{1}}^2=&
2 \lambda_{1} v_{1}^2
\, ,\\[+2mm] 
m_{H_{2}}^2=&
2 \lambda''_{10} v_{1}^2
\, ,\\ 
m_{H_{3}}^2=&
\frac{1}{2} \left(\lambda_{5} v_{1}^2+\lambda_{6}
   v_{2}^2+\lambda_{8} v_{1}^2+\lambda_{9} v_{2}^2+2
   \lambda''_{11} v_{1}^2-2 \lambda''_{12} v_{2}^2+2
   m^2_{33}\right)
\, ,\\ 
m_{A_{1}}^2=&
2 \lambda_{2} v_{2}^2
\, ,\\ 
m_{A_{2}}^2=&
2 \lambda''_{10} v_{2}^2
\, ,\\
m_{A_{3}}^2=&
\frac{1}{2} \left(\lambda_{5} v_{1}^2+\lambda_{6}
   v_{2}^2+\lambda_{8} v_{1}^2+\lambda_{9} v_{2}^2-2
   \lambda''_{11} v_{1}^2+2 \lambda''_{12} v_{2}^2+2
   m^2_{33}\right)
\, ,\\
m_{H^\pm_{1}}^2=&
-\frac{1}{2} (\lambda_{7}-2 \lambda''_{10})
   \left(v_{1}^2+v_{2}^2\right)
\, ,\\[+2mm]
m_{H^\pm_{2}}^2=&
\frac{1}{2} \left(\lambda_{5} v_{1}^2+\lambda_{6} v_{2}^2+2
   m^2_{33}\right)
\, .
\end{align}

\subsection{\texttt{F0CB}}

Looking at the results for the new minima found by our numerical simulation,
we realized that there is another particular case.
It is somewhat hidden because for $v_3=0$ we
are using too many angles. We identify a new situation, that we call
\texttt{F0CB}, and that can be defined by
\begin{equation}
 \label{eq:7}
  \sqrt{r_1}=\frac{v_1}{\sqrt{2}},
  \sqrt{r_2}=\frac{v_2}{\sqrt{2}},r_3=0,\quad
  \beta_1=\beta_2=\gamma=0,\quad
  \alpha_1\not=0,\alpha_2\not=0 .
\end{equation}
This is, in principle, a charge breaking minimum. We get the
stationary equations,
\begin{align}
 0=& \frac{1}{4} \left[4 m^2_{11} + 4 \lambda_{1} v_1^2 + 2 \lambda_{4} v_2^2
 + \lambda_{7} v_2^2 +  
 2 \lambda''_{10} v_2^2 + (\lambda_{7} + 2 \lambda''_{10}) v_2^2
 \cos(2 (\alpha_1 - \alpha_2))\right]\\ 
 0=&
 \frac{1}{4} \left[4 m^2_{22} + 2 \lambda_{4} v_1^2 + \lambda_{7} v_1^2 + 2 \lambda''_{10} v_1^2 + 
 4 \lambda_{2} v_2^2 + (\lambda_{7} + 2 \lambda''_{10}) v_1^2 \cos(2
 (\alpha_1 - \alpha_2))\right]\\ 
 0=& -\frac{1}{4} (\lambda_{7} + 2 \lambda''_{10}) v_1^2 v_2^2
 \sin(2 (\alpha_1 - \alpha_2))\\ 
 0=&\frac{1}{4} (\lambda_{7} + 2 \lambda''_{10}) v_1^2 v_2^2 \sin(2
 (\alpha_1 - \alpha_2)) 
\end{align}

These equations have many solutions.
As we showed analytically,
they are equivalent, giving the same value at the
minimum.
As an example, we take
\begin{align}
  \alpha_1=& \frac{\pi}{2}, \quad
  &&\alpha_2=0\\
  v_1^2=&-\frac{2 (2 \lambda_2 m^2_{11} - \lambda_4 m^2_{22})}{4 \lambda_1
    \lambda_2 - \lambda_4^2},\quad 
  &&v_2^2=-\frac{2 (-\lambda_4 m^2_{11} + 2 \lambda_1 m^2_{22})}{4 \lambda_1
    \lambda_2 - \lambda_4^2} 
\end{align}
and
\begin{equation}
  \label{eq:9}
  V_{\texttt{F0CB}}=-\frac{\lambda_{1} m^4_{22}+\lambda_{2}
      m^4_{11}-\lambda_{4} m^2_{11} 
   m^2_{22}}{4 \lambda_{1} \lambda_{2}-\lambda^2_{4}}
 \end{equation}

To have a local minimum, we take $v_1^2,v_2^2 >0$, as well as the
squared masses to be positive,
\begin{align} 
  m_{H_{1}}^2=&
\frac{4 \lambda_2 (\lambda_4 m^2_{11}-2 \lambda_1
   m^2_{22})}{4 \lambda_1 \lambda_2-\lambda_4^2}
\, ,\\[+2mm] 
m_{H_{2}}^2=&
\frac{(\lambda_7+2 \lambda''_{10}) (\lambda_4 m^2_{11}-2
   \lambda_1 m^2_{22})}{4 \lambda_1 \lambda_2-\lambda_4^2}
\, ,\\ 
m_{H_{3}}^2=&
\frac{1}{4 \lambda_1
  \lambda_2-\lambda_4^2}\left[
4 \lambda_1 \lambda_2 m^2_{33}-2 \lambda_1 \lambda_6
   m^2_{22}-2 \lambda_1 \lambda_9 m^2_{22}-4 \lambda_1
   \lambda''_{12} m^2_{22}-2 \lambda_2 \lambda_5
   m^2_{11}\right.\nonumber\\
 &\left. \hskip 20mm
   +\lambda_4^2 (-m^2_{33})
   +\lambda_4
   \lambda_5 m^2_{22}+\lambda_4 \lambda_6
   m^2_{11}+\lambda_4 \lambda_9 m^2_{11}+2 \lambda_4
   \lambda''_{12} m^2_{11} \right]
\, ,\\ 
m_{A_{1}}^2=&
\frac{(\lambda_7-2 \lambda''_{10}) (\lambda_4 m^2_{11}-2
   \lambda_1 m^2_{22})}{4 \lambda_1 \lambda_2-\lambda_4^2}
\, ,\\ 
m_{A_{2}}^2=&
\frac{1}{4 \lambda_1
  \lambda_2-\lambda_4^2}\left[
4 \lambda_1 \lambda_2 m^2_{33}-2 \lambda_1 \lambda_6
   m^2_{22}-2 \lambda_1 \lambda_9 m^2_{22}+4 \lambda_1
   \lambda''_{12} m^2_{22}-2 \lambda_2 \lambda_5
   m^2_{11}
\right.\nonumber\\
 &\left. \hskip 20mm
   +\lambda_4^2 (-m^2_{33})+\lambda_4
   \lambda_5 m^2_{22}+\lambda_4 \lambda_6
   m^2_{11}+\lambda_4 \lambda_9 m^2_{11}-2 \lambda_4
   \lambda''_{12} m^2_{11}\right]
\, ,\\
m_{H^\pm_{1}}^2=&
-\frac{2 \lambda_1 (2 \lambda_2 m^2_{11}-\lambda_4
   m^2_{22})}{4 \lambda_1 \lambda_2-\lambda_4^2}
\, ,\\[+2mm]
m_{H^\pm_{2}}^2=&
\frac{\lambda_4 \lambda_7 m^2_{22}-2 \lambda_2 \lambda_7
   m^2_{11}}{4 \lambda_1 \lambda_2-\lambda_4^2}
\, ,\\[+2mm]
m_{H^\pm_{3}}^2=&
\frac{1}{4
  \lambda_1 \lambda_2-\lambda_4^2}\left[
4 \lambda_1 \lambda_2 m^2_{33}-2 \lambda_1 \lambda_6
   m^2_{22}-2 \lambda_2 \lambda_5 m^2_{11}-2 \lambda_2
   \lambda_8 m^2_{11}
   \right.\nonumber\\
 &\left. \hskip 20mm
   +\lambda_4^2
   (-m^2_{33})+\lambda_4 \lambda_5 m^2_{22}+\lambda_4
   \lambda_6 m^2_{11}+\lambda_4 \lambda_8 m^2_{22}\right]
\, .
\end{align}

\section{Angles that minimize the angular part}
\label{app:f7f10_cond}

We have shown that the angular part of the potential
has a minimum when the functions $f_7$ and $f_{10}$ in Fig.~\ref{fig:f7Vsf10}
are themselves extrema. We have 7 equations, 3 for $f_7$ and 4 for $f_{10}$.
They are:
 \begin{eqnarray}
0 =   \frac{\partial f_7}{\partial \alpha_{+}} &=&
   2 (-1 + \cos\beta) \sin(2 \alpha_{+})\, ,
\label{D7alpha+}
   \\
0 =      \frac{\partial f_7}{\partial \alpha_{-}} &=&
   -2 (1 + \cos\beta) \sin(2 \alpha_{-})\, ,
\label{D7alpha-}
   \\
0 =      \frac{\partial f_7}{\partial \beta} &=&
  \left[\cos(2 \alpha_{+}) -\cos(2 \alpha_{-})\right] \sin\beta\, ,
\label{D7beta}
 \end{eqnarray}
and
\begin{eqnarray}
0 =      \frac{\partial f_{10}}{\partial \alpha_{+}} &=&
   -2 (-1 + \cos\beta) \cos(\beta - 2 \gamma) \sin(2 \alpha_{+}) + 
\nonumber\\
&& 4 \cos(\alpha_{-}) \sin(\alpha_{+}) \sin\beta \sin(\beta - 2 \gamma)\, ,
\label{D10alpha+}
   \\
0 =      \frac{\partial f_{10}}{\partial \alpha_{-}} &=&
   -2 (1 + \cos\beta) \cos(\beta - 2 \gamma) \sin(2 \alpha_{-}) +
\nonumber\\
&& 4 \cos(\alpha_{+}) \sin(\alpha_{-}) \sin\beta \sin(\beta - 2 \gamma)\, ,
\label{D10alpha-}
   \\
0 =   -   \frac{\partial f_{10}}{\partial \beta} - \frac{1}{2}  \frac{\partial f_{10}}{\partial \gamma}
&=& 
 4 \cos(\alpha_{-}) \cos(\alpha_{+}) \cos\beta \sin(\beta - 2 \gamma)  + 
   \nonumber\\
&& \left[ 2 + \cos(2 \alpha_{-}) + \cos(2 \alpha_{+}) \right]
\sin\beta \cos(\beta - 2 \gamma)\, ,
\label{aux6}
%
%
   \\
0 =     \frac{1}{2} \frac{\partial f_{10}}{\partial \gamma} &=&
   4 \cos(\alpha_{-}) \cos(\alpha_{+})  \sin\beta \cos(\beta - 2 \gamma) +
\nonumber\\
&&
\left[2 + \cos(2 \alpha_{-}) + \cos(2 \alpha_{+}) \right] \cos\beta \sin(\beta - 2 \gamma) +
   \nonumber\\
&&
\left[\cos(2 \alpha_{-}) - \cos(2 \alpha_{+}) \right] \sin(\beta - 2 \gamma)\, .
\label{D10gamma}
 \end{eqnarray}
We use Eq.~\eqref{aux6} instead of $\partial f_{10}/\partial \beta$
because it has a simpler form.
Eq.~\eqref{D7beta} provides two cases: $\sin\beta = 0$ and
$\cos(2 \alpha_{+}) = \cos(2 \alpha_{-})$, which we treat separately.

\subsection{\texorpdfstring{$\sin\beta = 0$}{sinbeta=0}; \texorpdfstring{$\beta = 0$}{}}

In this case, three amongst Eqs.~\eqref{D7alpha+}-\eqref{D10gamma} are automatically satisfied,
while the others become:
\begin{eqnarray}
0 &=& \sin{(2 \alpha_-)}\, ,
\nonumber\\
0 &=& \sin{(2 \alpha_-)} \cos{(2 \gamma)}\, ,
\nonumber\\
0 &=& \cos{(\alpha_-)} \cos{(\alpha_+)} \sin{(2 \gamma)}\, ,
\nonumber\\
0 &=& \cos^2{(\alpha_-)} \sin{(2 \gamma)}\, .
\label{Dwithbeta0}
\end{eqnarray}
This yields two cases. In the first
\begin{equation}
\beta = 0\, ,\ \ \  \cos\alpha_- = 0
\ \ \ \Longrightarrow \ \ \ 
f_7 = 0\, , \ \ \ f_{10} = 0\, .
\end{equation}
In the second case
\begin{equation}
\beta = 0\, ,\ \ \  \sin\alpha_- = 0\, ,\ \ \  \cos(2\gamma) = \pm 1
\ \ \ \Longrightarrow \ \ \ 
f_7 = 4\, , \ \ \ f_{10} = \pm 4\, .
\end{equation}

\subsection{\texorpdfstring{$\sin\beta = 0$}{sinbeta=0}; \texorpdfstring{$\beta = \pm \pi$}{}}

In this case, three amongst Eqs.~\eqref{D7alpha+}-\eqref{D10gamma} are automatically satisfied,
while the others become:
\begin{eqnarray}
0 &=& \sin{(2 \alpha_+)}\, ,
\nonumber\\
0 &=& \sin{(2 \alpha_+)} \cos{(2 \gamma)}\, ,
\nonumber\\
0 &=& \cos{(\alpha_-)} \cos{(\alpha_+)} \sin{(2 \gamma)}\, ,
\nonumber\\
0 &=& \cos^2{(\alpha_+)} \sin{(2 \gamma)}\, .
\end{eqnarray}
These are just Eqs.~\eqref{Dwithbeta0} with $\alpha_ - \leftrightarrow \alpha_+$.
Thus, we have again two cases.
In the first
\begin{equation}
\beta = \pi\, ,\ \ \  \cos\alpha_+ = 0
\ \ \ \Longrightarrow \ \ \ 
f_7 = 0\, , \ \ \ f_{10} = 0\, .
\end{equation}
In the second case
\begin{equation}
\beta = \pi\, ,\ \ \  \sin\alpha_+ = 0\, ,\ \ \  \cos(2\gamma) = \pm 1
\ \ \ \Longrightarrow \ \ \ 
f_7 = 4\, , \ \ \ f_{10} = \pm 4\, .
\end{equation}

\subsection{\texorpdfstring{$\sin\beta \neq  0$}{sinbeta!=0} and  \texorpdfstring{$\cos{(2 \alpha_+)}=\cos{(2 \alpha_-)}=1$}{}}

When $\sin{\beta} \neq 0$, Eqs.~\eqref{D7alpha+}-\eqref{D7alpha-} force $\sin(2\alpha_-)=0$ and
$\sin(2\alpha_-)=0$. Since Eq.~\eqref{D7beta} requires $\cos{(2 \alpha_+)}=\cos{(2 \alpha_-)}$,
we are left with the case $\cos{(2 \alpha_+)}=\cos{(2 \alpha_-)}=1$ treated here and
with the case  $\cos{(2 \alpha_+)}=\cos{(2 \alpha_-)}=-1$ treated in the next subsection.
Under the possibilities $(\alpha_+, \alpha_-) = (0,0), (\pi,\pi)$,
Eqs.~\eqref{aux6}-\eqref{D10gamma} force $\sin(2\beta-2\gamma)=0$.
And we find:
\begin{equation}
(\alpha_+, \alpha_-) = (0,0), (\pi,\pi)\, ,\ \ \  \cos(2\beta-2\gamma) = \pm 1
\ \ \ \Longrightarrow \ \ \ 
f_7 = 4\, , \ \ \ f_{10} = \pm 4\, .
\end{equation}
Similarly, under the possibilities $(\alpha_+, \alpha_-) = (0,\pi), (\pi,0)$,
Eqs.~\eqref{aux6}-\eqref{D10gamma} force $\sin(2\gamma)=0$,
and we find:
\begin{equation}
(\alpha_+, \alpha_-) = (0,\pi), (\pi,0)\, ,\ \ \  \cos(2\gamma) = \pm 1
\ \ \ \Longrightarrow \ \ \ 
f_7 = 4\, , \ \ \ f_{10} = \pm 4\, .
\end{equation}

\subsection{\texorpdfstring{$\sin{\beta} \neq 0$}{} and \texorpdfstring{$\cos{(2 \alpha_+)}=\cos{(2 \alpha_-)}=-1$}{}}

This is a very simple case, for all choices compatible with $\cos{(2 \alpha_+)}=\cos{(2 \alpha_-)}=-1$
yield:
\begin{equation}
\cos{(2 \alpha_+)}=\cos{(2 \alpha_-)}=-1
\ \ \ \Longrightarrow \ \ \ 
f_7 = 0\, , \ \ \ f_{10} = 0\, .
\end{equation}

\bibliographystyle{JHEP}
\bibliography{ref}

\end{document}